\documentclass[aip,jcp,preprint,superscriptaddress,floatfix]{revtex4-1}

\usepackage[utf8]{inputenc}
\usepackage{color}
\usepackage{graphicx}
\usepackage{mathtools}

\newcommand{\ket}[1]{\left|#1\right>}
\newcommand{\bra}[1]{\left<#1\right|}
\newcommand{\kPsi}{\ket{\Psi}}						
\newcommand{\ksig}[1][]{\ket{\sigma_{#1}}}			
\newcommand{\allStates}{\sigma_1,\ldots,\sigma_L}	
\newcommand{\allStatesopp}{\sigma_1^{\prime},\ldots,\sigma_L^{\prime}} 
\newcommand{\alldim}{a_1,\ldots,a_{L-1}}                     
\newcommand{\alldimop}{b_1,\ldots,b_{L-1}}                     
\newcommand{\ONvec}{\ket{\sigma_1 \ldots \sigma_L}}	
\newcommand{\ONvecb}{\bra{\sigma_1^{\prime} \ldots \sigma_L^{\prime}}}    
\newcommand{\ONstring}{\sigma_1 \ldots \sigma_L}		
\newcommand{\qcm}{\textsc{QCMaquis}}

\newcommand{\mR}{{\rm \textbf{R}}}
\newcommand{\mU}{{\rm \textbf{U}}}
\newcommand{\mT}{{\rm \textbf{T}}}
\newcommand{\mB}{{\rm \textbf{B}}}

\def\trio{[\textsc{trio}]$^{3-}$}

\begin{document}

\title[]{Second-Order Self-Consistent-Field Density-Matrix Renormalization Group} 

\author{Yingjin Ma}
 \email{yingjin.ma@phys.chem.ethz.ch}
\affiliation{%
ETH Z\"urich, Laboratorium f\"ur Physikalische Chemie, Vladimir-Prelog-Weg~2, 8093 Z\"urich, Switzerland
}%
\author{Stefan Knecht}
\email{stefan.knecht@phys.chem.ethz.ch}
\affiliation{%
ETH Z\"urich, Laboratorium f\"ur Physikalische Chemie, Vladimir-Prelog-Weg~2, 8093 Z\"urich, Switzerland
}%
\author{Sebastian Keller}
\affiliation{%
ETH Z\"urich, Laboratorium f\"ur Physikalische Chemie, Vladimir-Prelog-Weg~2, 8093 Z\"urich, Switzerland
}%
\author{Markus Reiher}
\email{markus.reiher@phys.chem.ethz.ch}
\affiliation{%
ETH Z\"urich, Laboratorium f\"ur Physikalische Chemie, Vladimir-Prelog-Weg~2, 8093 Z\"urich, Switzerland
}%

\date{\today}

\begin{abstract}
We present a matrix-product state (MPS)-based quadratically convergent density-matrix renormalization group self-consistent-field (DMRG-SCF) approach. 
Following a proposal by Werner and Knowles (\textit{J.~Chem.~Phys.~}\textbf{82}, 5053, (1985)), our DMRG-SCF algorithm is based on 
a direct minimization of an energy expression which is 
correct to second-order with respect to changes in the molecular orbital basis. 
We exploit a simultaneous optimization of the MPS wave function and molecular orbitals in order to achieve quadratic convergence. 
In contrast to previously reported (augmented Hessian) Newton-Raphson and super-configuration-interaction algorithms for DMRG-SCF, 
energy convergence beyond a quadratic scaling is possible in our \textit{ansatz}. Discarding the set of redundant active-active orbital rotations, 
the DMRG-SCF energy converges typically within two to four cycles of the self-consistent procedure. 
\end{abstract}

\maketitle

\section{Introduction}

Multiconfigurational self-consistent-field (MCSCF) theory \cite{szal11}\ including its most renowned complete-active space SCF (CASSCF) variant \cite{roos80,olse11} constitutes an integral part 
in the toolbox of modern quantum chemistry to describe the static part of the electron correlation energy \cite{roos08a}.  

Although molecular properties can sometimes be evaluated with sufficient accuracy based on an MCSCF-type approach, an 
MCSCF or CASCSF calculation will often only be a first step to obtain a reference configuration space and molecular orbital (MO) basis, 
which already takes into account important static correlation effects of the valence electrons. 
In a subsequent step, multi-reference approaches such as multi-reference configuration interaction (MRCI) or perturbation theory
(for example, complete active space second-order perturbation theory (CASPT2) \cite{ande90, ande92}), 
can be employed to recover dynamical electron correlation. 

In the CASSCF \textit{ansatz}, one defines an active space of $N$ electrons in $L$ orbitals, denoted as CAS($N$,$L$), 
in which we aim to find an exact solution of the electronic Schr{\"o}dinger equation by considering a full configuration interaction (FCI) expansion 
of the wave function while simultaneously optimizing the \textit{full} MO basis. 
Suffering from an exponential growth\cite{aqui15a} of the FCI expansion with respect to increasing values of $N$ and $L$, 
active orbital spaces beyond CAS(18,18) are out of reach for CASSCF based on traditional configuration interaction (CI) expansions.

By contrast, the density-matrix renormalization group (DMRG) approach \cite{whit92,whit93,scho05,scho11} in quantum chemistry
\cite{lege08,chan08,chan09,mart10,mart11,chan11a,wout14,kura14a,yana15,oliv15a,szal15,knec16a,chan16}\ is capable of approximating CAS-type wave functions 
to chemical accuracy with polynomial scaling. In combination with a self-consistent-field orbital optimization ansatz (DMRG-SCF) \cite{zgid08,ghos08,zgid08b},  
active orbital spaces of about five to six times the CASSCF limit are accessible. The selection of a suitable active orbital space is a 
tedious procedure, but may be automatized\cite{stei16a,stei16b}.

Given a pre-defined active orbital space, present DMRG-SCF algortihms can be divided up into two conceptually different approaches. 
The first approach \cite{zgid08, knec16a} exploits the generalized Brillouin theorem \cite{levy68} where orbital changes are obtained from the 
coefficients of a so-called 'Super-CI' procedure \cite{grei71, bane77, roos80} consisting of 
the DMRG-SCF wave function and all Brillouin singly excited configurations. 
Such an orbital-optimization scheme is implemented in many popular quantum chemical packages, 
for example, in {\sc Molcas} \cite{aqui15a} and {\sc Orca} \cite{nees12o}.
The convergence behavior of the Super-CI algorithm is quite robust although 
it is not a quadratically convergent algorithm \cite{chan77, shep87}.

The second class of DMRG-SCF approaches focuses on a direct minimization of the energy. 
This \textit{ansatz}\ was first exploited for DMRG-SCF in the [augmented Hessian (AH)] Newton-Raphson-like (NR) 
implementation by Ghosh and co-workers \cite{ghos08} and Wouters \textit{et al.} \cite{wout14b,wout15}. Its implementation was also 
described by Ma and Ma \cite{mayi13}\ who, in addition, presented a pilot DMRG-SCF implementation of the Werner-Meyer (WM) MCSCF algorithm \cite{wern80}.
Their common basis is the construction of operators which are obtained from 
taking the first and second derivatives of the energy with respect to the variational parameters. 

In the NR algorithm and its AH form\cite{leng80}, the energy expression is chosen to be a second-order function 
of the orbital rotation parameters and the changes in the wave function expansion coefficients\cite{yeag82}. 
By contrast, the WM algorithm employs a second-order energy expression which is periodic in the orbital changes and contains higher-order terms in 
the orbitals rotation parameters\cite{wern80,wern85}. This ensures a large convergence radius even for ill-conditioned 
starting configurations, where the Hessian (the matrix of second derivatives with respect to the variational parameters) 
exhibits a large number of negative eigenvalues\cite{wern85,know85}. 

A distinct feature of the energy-based orbital-optimization algorithms is the possibility to take into account coupling terms 
between the single- and many-particle-basis parameters to simultaneously update the wave function and orbitals \cite{shep87, wern87, szal11}.
Their calculation requires for either approach, NR, AH, or WM, access to derivatives of the one- and two-particle 
reduced density matrices (RDMs), i.e., symmetrized transition density matrices. 
In this work, we employ a matrix product state (MPS) and matrix product operator (MPO) formalism which 
allows us to formulate the calculation of the RDM derivatives in a straightforward fashion. With these quantities at hand, 
we outline briefly how to derive the coupling terms in the Hessian for a simultaneous update of the wave function and 
orbital parameters \cite{shep87, szal11}. By contrast, as proposed by Werner and Knowles \cite{wern85}, it is possible to entirely 
avoid the calculation of the RDM derivatives by introducing the coupling for the 
WM-based optimization algorithm through solving a set of coupled nonlinear equations 
employing a second-order Hamiltonian expression \cite{wern85, wern87}. We therefore preferred to pursue this approach while 
studying different approximations to obtain a relaxed MPS with respect to changes in the MO basis.

This paper is organized as follows: on the basis of the work of Werner, Meyer, and Knowles (WMK)\cite{wern80,wern85,know85}, 
the central element of our algorithm is the coupled optimization of the MPS and MOs
to obtain a quadratically convergent DMRG-SCF approach.  In Section \ref{sec:wmk}, we outline our orbital-optimization approach 
for MPS wave functions and discuss possibilities for a simultaneous optimization of the MPS(s) and MOs in 
Section \ref{sec:simulOPT}). Since one of these coupling approaches requires an evaluation of the first derivative 
of the one- and two-particle RDMs with respect to the variational parameters, we illustrate their derivation within 
the MPS framework in Section \ref{sec:RDMder}. Numerical examples are presented in Section \ref{sec:numerical}.

\section{Theory}\label{sec:method}

\subsection{Orbital Optimization for Matrix Product State Wave Functions}\label{sec:wmk}

In this section, we briefly summarize the orbital optimzation algorithm for multiconfigurational wave functions as proposed by Werner, Meyer, and Knowles. 
A detailed summary of the formulation based on CI-type wave functions can be found in their original papers, Refs. \cite{wern80,wern85,know85}. In addition, a comprehensive review by Werner~\cite{wern87} is available. 
Hence, we restrict our discussion in this work to the essential steps necessary to illustrate our formulation of the WMK optimization scheme within the framework of MPS multiconfigurational wave functions. 
In what follows, (doubly- and partially) occupied orbitals will be labeled by {\it i, j, k, l,}\ whereas {\it r, s, t} will denote arbitrary (occupied or empty) MOs. We further assume that the wave function under consideration is real which allows us to restrict the orbitals to be real. In addition, we work in a spin-restricted formalism such that each orbital can be occupied by up to two electrons with opposite spin. 

\subsubsection{Matrix Product State Wave Functions and Matrix Product Operators}

In a traditional CI \textit{ansatz}, we can express an arbitrary state $\kPsi$\ in a Hilbert space spanned 
by $L$\ spatial orbitals as a linear superposition of occupation number vectors $\ket{\boldsymbol{\sigma}}$\ 
with the CI coefficients $c_{\ONstring}$\ as expansion coefficients, 
\begin{equation}\label{eq:CI_wave_function}
\kPsi = \sum\limits_{{\boldsymbol{\sigma}}} c_{\boldsymbol{\sigma}} \ket{\boldsymbol{\sigma}} = \sum\limits_{\allStates} c_{\ONstring} \ONvec\ .
\end{equation}
Each local space is of dimension four corresponding to the 
basis states $\sigma_l = \left|\uparrow\downarrow\right>, \left|\uparrow\right>, \left|\downarrow\right>, \left|0\right>$\ 
of the $l$-th spatial orbital. 
Turning to an MPS representation of $\kPsi$, we encode the CI 
coefficients $c_{\ONstring}$\ as a product of $m_{l-1}\times m_{l}$-dimensional matrices 
$M^{\sigma_l} = \{M^{\sigma_l}_{a_{l-1}a_l}\}$ 
\begin{align}
\kPsi &= \sum_{\allStates} \sum_{\alldim} M^{\sigma_1}_{1 a_1} M^{\sigma_2}_{a_1 a_2} \cdots M^{\sigma_L}_{a_{L-1} 1} \ONvec = \sum_{\boldsymbol{\sigma}} M^{\sigma_1} M^{\sigma_2} \cdots M^{\sigma_L} \ket{\boldsymbol{\sigma}}\ ,\label{eq:MPS2}
\end{align}
where the last equality is a result of collapsing the summation over 
the $a_l$\ indices (typically referred to as \emph{virtual indices} or \emph{bonds}) as matrix-matrix multiplications. 
Note that the first and the last matrices are $1\times m_1$-dimensional row and $m_{L-1}\times 1$-dimensional column vectors, respectively, 
since the final contraction of the matrices $M^{\sigma_l}$\ must yield the scalar coefficient $c_{\ONstring}$. 

The central idea that facilitates a reduction of the exponentially scaling full CI \textit{ansatz} in Eq.~(\ref{eq:CI_wave_function}) 
to a polynomial-scaling MPS wave function \textit{ansatz}\ is the introduction of some maximum dimension $m$ for the matrices $M^{\sigma_l}$, 
where $m$\ is the \textit{number of renormalized block states} \cite{whit93}. 
We refer the reader for further details on the actual variational search algorithm for ground- and excited states in an MPS framework 
to the review by Schollw\"ock \cite{scho11}\ and to our recent papers \cite{kell15a,kell16} for a detailed description in our implementation. 

In passing, we note that we may further exploit the matrix-product formulation 
to express an operator $\widehat{W}$\ in MPO form \cite{scho11}
\begin{eqnarray}\label{eq:MPO}
\widehat{\mathcal{W}} & = & \sum_{\allStates} \sum_{\allStatesopp} \sum_{\alldimop} W^{\sigma_1 \sigma_1^{\prime}}_{1 b_1} W^{\sigma_2 \sigma_2^{\prime}}_{b_1 b_2} \cdots W^{\sigma_L \sigma_L^{\prime}}_{b_{L-1} 1} 
\ONvec \ONvecb\nonumber\\
& = & \sum_{\boldsymbol{\sigma},\boldsymbol{\sigma^{\prime}}} W^{\sigma_1 \sigma_1^{\prime}} W^{\sigma_2 \sigma_2^{\prime}} \cdots W^{\sigma_L \sigma_L^{\prime}}  \ket{\boldsymbol{\sigma}}\bra{\boldsymbol{\sigma^{\prime}}} \equiv \sum_{\boldsymbol{\sigma},\boldsymbol{\sigma^{\prime}}} w_{\boldsymbol{\sigma} \boldsymbol{\sigma^{\prime}}} \ket{\boldsymbol{\sigma}}\bra{\boldsymbol{\sigma^{\prime}}}\ ,
\end{eqnarray}
with incoming and outgoing physical states $\sigma_l$\ and $\sigma_l^{\prime}$\, respectively, and the virtual indices $b_{l-1}$\ and $b_l$. 
In complete analogy to Eq.~(\ref{eq:MPS2}), the summation over pairwise matching indices $b_l$\ can be regarded  
as matrix-matrix multiplications which yields the second line on the right-hand side of Eq.~(\ref{eq:MPO}). 
To be of practical use, the summations in Eq.~({\ref{eq:MPO}})\ are rearranged such that the 
contraction is carried out first over the local site indices $\sigma_l, \sigma_l^{\prime}$\ 
\begin{equation}\label{eq:local_contract}
\widehat{W}^{l}_{b_{l-1} b_{l}} = \sum\limits_{\sigma_l,\sigma_l^{\prime}} W^{\sigma_l \sigma_l^{\prime}}_{b_{l-1} b_l} \ket{\sigma_l} \bra{\sigma_l^{\prime}}\ ,
\end{equation}
which leads to 
\begin{equation}\label{eq:mpoMOD}
\widehat{\mathcal{W}}  = \sum\limits_{\alldimop} \widehat{W}^{1}_{1 b_{1}} \cdots \widehat{W}^{l}_{b_{l-1} b_{l}} \cdots \widehat{W}^{L}_{b_{L-1} 1}\ = 
\widehat{W}^{1} \cdots \widehat{W}^{l} \cdots \widehat{W}^{L}\ .
\end{equation}
In Eq.~(\ref{eq:local_contract}), we introduced a local, operator-valued matrix representation $\widehat{W}^{l}_{b_{l-1} b_{l}}$\ 
which is a key element for an efficient MPO-based implementation of the quantum-chemical 
DMRG approach \cite{kell15a,kell16}\ that offers the same polynomial scaling as a 'traditional' (non-MPO) DMRG implementation. 

As a result of the rearrangement, the entries of the $\widehat{W}^{l}_{b_{l-1} b_{l}}$\ matrices 
comprise the elementary, \textit{local} operators acting on the $l$-th orbital such as for example the creation and annihilation operators 
$\hat{c}_{\tau_l}^{\dagger}$\ and  $\hat{c}_{\tau_l}^{}$. In order to illustrate this point,
we express the operator $\hat{c}_{\uparrow_l}^{\dagger}$\ as a linear combination of the local basis states 
\begin{equation}\label{eq:create-sum}
\hat{c}_{\uparrow_l}^{\dagger} = \left|\uparrow \downarrow\right>\left<\downarrow\right| + \left|\uparrow\right>\left<0\right|\ .
\end{equation}
From Eq.~(\ref{eq:create-sum}) it follows that the corresponding matrix representation for the operator $\hat{c}_{\uparrow_l}^{\dagger}$\ 
is a $(4\times4)$-dimensional matrix with two non-zero entries equal to one. Similar considerations hold for the remaining local operators. 
Consequently, the MPO formulation allows us to efficiently arrange the creation and annihilation operators of the 
full (non-relativistic) electronic Hamiltonian $\hat{H}$, 
\begin{equation}\label{Hami_op}
\hat{H} = \sum_{i,j,\tau}^{} (i |h| j)\ \hat{c}_{i\tau}^{\dagger} \hat{c}_{j\tau}+ \frac{1}{2} \sum\limits_{\substack{i,j,k,l\\ \tau,\tau^{\prime}}}^{} (ij|kl)\ \hat{c}_{i\tau}^{\dagger} \hat{c}_{k\tau^{\prime}}^{\dagger} \hat{c}_{l\tau^{\prime}} \hat{c}_{j\tau}\ ,
\end{equation}
into the operator valued matrices introduced in Eq.~(\ref{eq:mpoMOD}). For details on how to achieve the latter in an efficient way, 
we refer the reader to Ref.~\citenum{kell15a}. The parameters
\begin{equation}\label{1eINT}
(i |h| j) = \int \phi_i^*({\rm \textbf{r}}) \hat{h} \phi_j({\rm \textbf{r}})d{\rm \textbf{r}}
\end{equation}
and 
\begin{equation}\label{2eINT}
(ij|kl) = \int \phi_i^*({\rm \textbf{r}}_1)\phi_k^*({\rm \textbf{r}}_2) r_{12}^{-1} \phi_j({\rm \textbf{r}}_1)\phi_l({\rm \textbf{r}}_2)d{\rm \textbf{r}}_1 d{\rm \textbf{r}}_2
\end{equation}
are the one- and two-electron integrals in the orthonormal MO basis $\{\phi_l\}$,  i.e., $\left<k\left.\right|l\right> = \delta_{kl}$\ 
holds for all $k$\ and $l$. The MOs $\{\phi_l\}$ are represented as linear combinations of 
the atomic orbital basis functions {$\{\chi_{\mu}\}$} 
\begin{equation}\label{orbMOAO}
\phi_l = \ket{l} = \sum_{\mu} C_{\mu l} \chi_{\mu}\ ,
\end{equation}
where $\{C_{\mu l}\}$ are the molecular orbital coefficients. 

\subsubsection{Orbital Optimization}

Given one (\textit{state-specific}) or several (\textit{state-averaged}) MPS wave function(s)
of the form in Eq.~(\ref{eq:MPS2}) and expressed in the MO basis $\{\phi_l\}$,  the objective of DMRG-SCF is 
to minimize the energy given by 
\begin{equation}\label{energy}
 E_0  = \sum_{i,j} \left<i | h | j\right> \gamma_{ij} + \frac{1}{2} \sum_{i,j,k,l} (ij|kl) \Gamma_{ijkl}\ , 
\end{equation}
with respect to the MPS parameters $\{M^{\sigma_l}\}$\ and the molecular orbitals $\{\phi_l\}$\ which constitute the basis for the occupation number vectors $\boldsymbol{\ksig}$.
Here, 
\begin{equation}\label{1eRDM}
\gamma_{ij}  = \sum_{\tau}\left<\Psi\right|\hat{c}_{i\tau}^{\dagger}\hat{c}_{j\tau}\left|\Psi\right>
\end{equation}
and 
\begin{equation}\label{2eRDM}
\Gamma_{ijkl} = \sum_{\tau,\tau^{\prime}}\left<\Psi\right|\hat{c}_{i\tau}^{\dagger}\hat{c}_{k\tau^{\prime}}^{\dagger}\hat{c}_{l\tau^{\prime}}^{}\hat{c}_{j\tau}\left|\Psi\right>
\end{equation}
are elements of the one-and two-particle RDMs, respectively.

Invoking the variation principle, a new set of orthonormal MOs $\{\tilde{\phi}_l^{}\}$, 
which will minimize the energy expectation value given in Eq.~(\ref{energy}), can be obtained from a unitary transformation 
of the orbitals 
\begin{equation}\label{neworb}
\ket{\tilde{i}^{}} = \sum_r \ket{r} U_{ri}\ ,
\end{equation}
where \textbf{U}=$\{U_{ri}\}$\ can be expressed in exponential form \cite{thou61,levy69a}
\begin{equation}\label{eR}
 {\rm \textbf{U}(\textbf{R})} = e^{\rm \textbf{R}} =1 +  {\rm \textbf{R}} + \frac{1}{2} {\rm \textbf{R}}{\rm \textbf{R}} + \cdots\ .
\end{equation}
\textbf{R} is an antihermitean matrix (i.e. ${\rm -\textbf{R}} = {\rm \textbf{R}} ^{\dagger}$)
comprising a set of independent orbital rotation parameters $\{R_{ri}\}$ with $r > i$. Moreover, the energy expectation value in
Eq.~(\ref{energy}) is also a function of the MPS parameters from which the density matrices $\gamma_{ij}$\ and $\Gamma_{ijkl}$ 
are calculated. This dependence will be discussed in Section \ref{sec:simulOPT}.  

By setting $R_{rs} = 0$,  where 
$r$\ and $s$ refer to doubly occupied (inactive) or empty (secondary) orbitals, redundant parameters 
which do not affect the energy to first oder can be removed. Because of the two-electron terms, 
the energy expression Eq.~(\ref{energy}) is a fourth-order function of \textbf{U}\ which makes a direct optimization with respect to \textbf{U}\ impractical. 
Following Werner and Meyer \cite{wern80}, by introducing an auxiliary matrix \textbf{T}\ 
\begin{equation}\label{UT}
{\rm \textbf{T}} = {\rm \textbf{U}} - 1 = {\rm \textbf{R}} + \frac{1}{2} {\rm \textbf{R}\textbf{R}} + \cdots\ ,
\end{equation}
and expanding the energy functional in Eq.~(\ref{energy}) up to order $\mathcal{O}(T^2)$\ we obtain an energy expression of the form\cite{wern85,wern87}
\begin{align}
\label{E2origin}
E^{(2)}(\mT) = &\ E_0 +  2  \sum_{r,i} T_{ri} \left[\sum_{j}\left<r\left|h\right|j\right> \gamma_{ij} + \sum_{j,k,l} \left<r\left|J^{kl}\right|j\right>\Gamma_{ijkl}\right]  \nonumber\\
& +\sum_{r,i} \sum_{s,j} T_{ri} T_{sj} \left[\left<r\left|h\right|s\right> \gamma_{ij} + \sum_{k,l} \bigg(\left<r\left|J^{kl}\right|s\right>\Gamma_{ijkl} 
+ 2 \left<r\left|K^{kl}\right|s\right>\Gamma_{ikjl}\bigg)\right]\ \nonumber\\
= & \ E_0 + 2 \sum_{r,i} T_{ri} A_{ri} +  \sum_{i,j,r,s} T_{ri} \left< r| G^{ij} | s \right> T_{sj} \nonumber \\
= &\ E_0 + \sum_{r,i} T_{ri} \left(A_{ri} + B_{ri}\right)\ ,
\end{align}
with the operator $G^{ij}$\ 
\begin{equation}\label{opG}
G^{ij} = h\gamma_{ij}\ +\ \sum_{k,l} \Gamma_{ijkl} J^{kl}\ +\ 2\sum_{k,l} \Gamma_{ikjl} K^{kl}
\end{equation}
and the matrices 
\begin{align}\label{A}
A_{ri} =&\ \sum_{j}\left<r\left|h\right|j\right> \gamma_{ij}\ +\ \sum_{j,k,l} \left<r\left|J^{kl}\right|j\right>\Gamma_{ijkl} 
\end{align}
and
\begin{align}\label{Borig}
B_{ri} =&\ A_{ri} + \sum_{s,j} \left<r\left|h\right|s\right> \gamma_{ij} T_{sj} +\  \sum_{k,l} \sum_{s,j} \bigg(\left<r\left|J^{kl}\right|s\right>\Gamma_{ijkl} +\ 2 \left<r\left|K^{kl}\right|s\right>\Gamma_{ikjl} \bigg) T_{sj}\ \nonumber \\
 =&\ \sum_{s,j}\left<r\left|h\right|s\right> \gamma_{ij} U_{sj} + \sum_{k,l} \sum_{s,j} \left<r\left|J^{kl}\right|s\right>\Gamma_{ijkl} U_{sj} +\ 2 \left<r\left|K^{kl}\right|s\right>\Gamma_{ikjl} T_{sj}\ .
\end{align}
In passing, we note that \mB\ = $\{B_{ri}\}$\ depends on \mU\ (and therefore on \mT\ = \mU\ $-$ {\bf 1}), which will be considered further below. 
The generalized Coulomb ${J}^{kl}$\ and exchange operators $K^{kl}$\ are defined through 
\begin{equation}\label{opJ}
\left<r\left|J^{kl}\right|s\right> = \left(rs|kl\right)
\end{equation}
and 
\begin{equation}\label{opK}
\left<r\left|K^{kl}\right|s\right> = \left(rk|ls\right)\ .
\end{equation}
Eq.~(\ref{E2origin})\ will be minimized iteratively (\textit{macro iterations, vide infra}) 
until self-consistency is reached, i.e., when all orbital changes are smaller than a given threshold. 
An extension of the energy expansion in Eq.~(\ref{E2origin})\ with respect to rotations of inactive orbitals is straightforward and the respective formulae can be found in Ref.~\citenum{wern80}, Eqs.~(62)-(70).
Moreover, as can be seen from the terms appearing in Eq.~(\ref{E2origin}), their evaluation requires, in accord with other 
second-order MCSCF schemes, at each macro-iteration step an integral transformation to the MO basis of the current expansion point 
with at most two general indices $r,s$. In contrast to the Super-CI algorithm \cite{roos80} and NR \cite{sieg81, jorg81,yeag82} (or AH \cite{leng81}) approaches, 
which set out from a truncation of the series expansion in Eq.~\eqref{eR} to first and second order in \mR 
resulting in a second-order energy approximation $E^{(2)}$\ as a function of \mR, Eq.~(\ref{E2origin}) contains terms of infinite order in \mR\ (cf. Eq.~(\ref{UT})). This feature was shown to significantly improve both the radius of convergence and convergence properties of the WMK scheme compared to the aforementioned approaches, in particular at expansion points far from the (local) minimum, see Refs.~\citenum{wern85} and \citenum{wern87}\ for further details.

Following Werner and Knowles\cite{wern85,know85,wern87}, we define an update of \mU(\mR)\ as a multiplication of \mU(\mR)\ with an additional unitary matrix \mU($\boldsymbol{\Delta}$\mR),
\begin{align}\label{updateU}
{\rm \mU}({\rm \mR},\boldsymbol{\Delta}{\rm \mR}) =&\ {\rm \mU}({\rm \mR}) \cdot {\rm \mU}(\boldsymbol{\Delta}{\rm \mR})\ \nonumber \\
=&\ {\rm \mU} + {\rm \mU} \cdot (\boldsymbol{\Delta}{\rm \mR} + \frac{1}{2}\boldsymbol{\Delta}{\rm \mR}\boldsymbol{\Delta}{\rm \mR} + \ldots)\ ,
\end{align}
where the antisymmetric matrix $\boldsymbol{\Delta}\mR = - \boldsymbol{\Delta}\mR^{\dagger}$\ describes the change of \mU.
Equivalently for $\boldsymbol{\Delta}{\rm \mT}$\ we obtain,
\begin{align}\label{updateT}
\boldsymbol{\Delta}{\rm \mT} =&\ {\rm \mU} \cdot (\boldsymbol{\Delta}{\rm \mR} + \frac{1}{2}\boldsymbol{\Delta}{\rm \mR}\boldsymbol{\Delta}{\rm \mR} + \ldots)\ ,
\end{align}
with \mT(\mR,$\boldsymbol{\Delta}$\mR) = \mT + $\boldsymbol{\Delta}$ \mT\ and the definitions given in Eqs.~(\ref{UT}) and (\ref{updateU}). 
Inserting Eq.~(\ref{updateU})\ into Eq.~(\ref{E2origin}) and expanding the energy approximation Eq.~(\ref{E2origin})\ up to second order in $\boldsymbol{\Delta}\mR$ yields for a given \mU\ \cite{wern85,know85,wern87}
 \begin{align}\label{E2deltaR}
 E^{(2)}(\mT,\boldsymbol{\Delta}{\rm \mR}) = &\ E^{(2)}(\mT) +\ 2 \sum_{r,i} \big(\Delta R_{ri} + \frac{1}{2}(\boldsymbol{\Delta}\mR)^{2}_{ri}\big) \tilde{A}_{ri} + \sum_{s,i,t,j} \Delta R_{si} (\mU^{\dagger}\textbf{G}^{ij}\mU)_{st} \Delta R_{tj}\ ,
 \end{align}
 with
 \begin{align}\label{Aupdate}
 \tilde{A}_{ri} =&\ (\mU^{\dagger}\mB)_{ri}\ ,
 \end{align}
 and
 \begin{equation}\label{opGupdate}
 (\textbf{G}^{ij})_{st} = \left<s\left|h\right|t\right> \gamma_{ij}\ +\ \sum_{k,l} \left<s\left|J^{kl}\right|t\right> \Gamma_{ijkl} +\ 2 \sum_{k,l} \left<s\left|K^{kl}\right|t\right> \Gamma_{ikjl} \ .
 \end{equation}
 
It can then be shown \cite{wern80,know85,wern85,wern87} that the resulting energy approximation $E^{(2)}(\mT,\boldsymbol{\Delta}{\rm \mR})$ has a stationary point with respect to variations of \mT\ if the following 
conditions are satisfied
 \begin{equation}\label{UBBU0}
\left(\frac{\partial E^{(2)}(\mT,\boldsymbol{\Delta}{\rm \mR})}{\partial \Delta R_{ri}}\right)_{\boldsymbol{\Delta}{\rm \mR} = 0}\ = 2(\tilde{\rm \textbf{A}} - \tilde{\rm \textbf{A}}^{\dagger})_{ri} = 0\quad \forall\ r > i\ ,
 \end{equation}
which, by inserting Eq.~(\ref{Aupdate}), takes the form of a nonlinear matrix equation \cite{wern80,wern85,know85,wern87},
 \begin{equation}\label{UBBU}
 \mU^{\dagger}\mB-\mB^{\dagger}\mU  = 0\ .
 \end{equation}
 Recalling the dependence of \mB\ on \mU\ (cf. Eq.~(\ref{Borig})), the nonlinear Eqs.\ (\ref{UBBU})\ are best solved 
 iteratively to determine the optimal \mU. Each of these steps, usually denoted as \textit{micro iterations}, requires a new 
 evaluation of \mB. This comprises a one-index transformation of $\textbf{h}$, $\textbf{J}^{kl}$, and $\textbf{K}^{kl}$\ \cite{wern85,know85,wern87}
 \begin{align}
(\tilde{\textbf{h}})_{rj} =&\ ({\textbf{h}}\mU)_{rj}\ , \label{eq:halfh} \\
(\tilde{\textbf{J}}^{kl})_{rj} =&\ ({\textbf{J}}^{kl}\mU)_{rj}\ , \label{eq:halfJ} \\
(\tilde{\textbf{K}}^{kl})_{rj} =&\ ({\textbf{K}}^{kl}\mU)_{rj}\ \label{eq:halfK} .
 \end{align}
Once the micro iterations are converged, i.e., Eq.~(\ref{UBBU})\ is fulfilled by \mU\ to a preset accuracy, a new set of MOs is obtained from the final \mU\ according to Eq.~(\ref{neworb}). This step marks then the beginning of the next \textit{macro-iteration}\ step, 
which in turn sets out from an evaluation of a new set of $\textbf{h}$, $\textbf{J}^{kl}$, $\textbf{K}^{kl}$\ as well as of
the expectation value $E^{(2)}(\mT)$ (Eq.~(\ref{E2origin})). The macro iterations will be converged if the lowering in $E^{(2)}(\mT)$\ between two 
macro-iteration steps, $\delta E$, becomes smaller than a preset threshold. 

In order to solve Eq.~(\ref{UBBU}) (for a fixed set of $M^{\sigma_l}$\ matrices in an MPS representation of $\kPsi$) we 
adopt the 'direct-MCSCF' idea of Werner and Knowles \cite{wern85,know85,wern87}\ and employ a step-restricted augmented Hessian approach to minimize the second-order energy approximation $E^{(2)}(\mT,\boldsymbol{\Delta}{\rm \mR})$,
\begin{equation}\label{eq:augHess}
\left(
\begin{array}{cc}
-\epsilon & \textbf{g}^{\dagger} \\
\textbf{g} & \textbf{H}/\lambda - \epsilon
\end{array}
\right)
\left(
\begin{array}{c}
1/\lambda \\
\textbf{x}
\end{array}
\right) = 0\ ,
\end{equation}
with
\begin{equation}\label{eq:eigvalAH}
\epsilon =  \lambda \textbf{g}^{\dagger} \textbf{x}\ ,
\end{equation}
and \textbf{x} = $\{\Delta R_{ri}\}$\ with $r > i$. While solving the above linear Eq.\ (\ref{eq:eigvalAH}) employing a Davidson-type approach \cite{davi75}, the damping parameter $\lambda$\ is successively adjusted in an automatized fashion to ensure that the step length $|\textbf{x}|$\ remains less than
or equal to a predefined maximum step length $s$, 
by, for example, requiring $\sum\limits_{r,i} \Delta R_{ri} \leq\ s^2$. With the elements of the gradient \textbf{g}\ given by Eq.~(\ref{UBBU0})\ and the elements of the Hessian \textbf{H}\ by \cite{wern87}
\begin{equation}\label{ETdRH}
\left( \frac{\partial E^{(2)} (\mT,\boldsymbol{\Delta}{\rm \mR})}{\partial \Delta R_{ri} \partial \Delta R_{sj}} \right)_{\Delta R =0}
= (1-\tau_{ri})(1-\tau_{sj}) \left(2(\mU^{\dagger}\textbf{G}^{ij}\mU)_{rs} - (\tilde{A}_{rs}+\tilde{A}^{\dagger}_{rs})\delta_{ij}\right)\ ,
\end{equation}
where $\tau_{ri}$\ permutes the indices $r$\ and $i$, 
the residual vector \textbf{y},
\begin{equation}\label{eq:y}
\textbf{y} = \textbf{g} + (\textbf{H} - \lambda\epsilon)\textbf{x}\ ,
\end{equation}
which needs to be evaluated iteratively in the Davidson diagonalization approach, can be expressed in matrix form as \cite{wern85,know85,wern87}
\begin{equation}\label{eq:Ymat}
\textbf{Y} =  2(\mU^{\dagger}\tilde{\mB}-\tilde{\mB}^{\dagger}\mU) - (\tilde{\textbf{A}} - \tilde{\textbf{A}}^{\dagger}) \boldsymbol{\Delta}{\rm \mR} + \boldsymbol{\Delta}{\rm \mR}(\tilde{\textbf{A}} - \tilde{\textbf{A}}^{\dagger}) - \lambda \epsilon \boldsymbol{\Delta}{\rm \mR}\ .
\end{equation}
The elements of $\tilde{\mB}$ are defined as \cite{wern85,know85,wern87}
\begin{equation}\label{eq:Bupdate}
\tilde{B}_{ri} = B_{ri} + \sum_j (\textbf{G}^{ij}\mU\boldsymbol{\Delta}{\rm \mR})_{rj}\ ,
\end{equation}
and 
\begin{equation}\label{eq:epsDAV}
\epsilon = 2 \lambda {\rm tr}(\tilde{\textbf{A}}^{\dagger}\boldsymbol{\Delta}{\rm \mR})\ ,
\end{equation}
where 'tr' denotes the trace of a matrix. 

After convergence of the Davidson procedure corresponding to the residual \textbf{Y}\ becoming smaller than a preset threshold, the solution $\boldsymbol{\Delta}{\rm \mR}$\ is used to update \mU\ (cf.~Eq.~(\ref{updateU})). This is followed by an evaluation of \mB\ which initializes a 
new micro iteration step until Eq.~(\ref{UBBU})\ is fulfilled to a desired accuracy. 
For further details, in particular concerning the Davidson procedure outlined above, we refer to Ref.~\citenum{wern87}. 
We note that for \textbf{T}\ = 0\ the gradient and Hessian expressions, Eqs.~(\ref{UBBU0}) and (\ref{ETdRH}), reduce to the derivative expressions encountered for a conventional AH approach \cite{wern87}\ which we exploited in the present work 
for our implementation of the AH approach for DMRG-SCF \cite{zgid08, wout15, yana15}. 

As indicated above, the iterative approach to solve Eq.~(\ref{UBBU})\ does so far not take 
into account any relaxation of the MPS parameters $\{M^{\sigma_l}\}$\ and neglecting the latter will result in a loss of 
quadratic convergence which can be seen for the numerical examples presented in Section \ref{sec:numerical}. 
That the solution of Eq.~(\ref{UBBU})\ \textit{is} coupled to a relaxation of the MPS parameters can easily be seen from the dependence of \mB\ in Eq.~(\ref{Borig})\ on the one- and two-particle RDMs which are calculated from the MPS $\kPsi$\ at the current expansion point (cf.~Eqs.(\ref{1eRDM}) and (\ref{2eRDM})). Hence, as will be discussed in the next section, we achieve a coupling 
in the micro iterations\ by adapting the procedure proposed by Werner and Knowles \cite{wern85,know85,wern87} 
for conventional CI wave functions to our  MPS framework and carry out a direct-CI-type step each time \mU\ is updated 
according to Eq.~(\ref{updateU}). The relaxed MPS wave function is then employed to evaluate new RDMs which 
in turn are needed for the calculation of \mB.

\subsection{Simultaneous optimization of MPS parameters and orbitals}\label{sec:simulOPT}

\subsubsection{Concepts}\label{sec:concepts}

Within the two-site DMRG optimization algorithm which we employ in the present work, 
the MPS wave function of Eq. \eqref{eq:MPS2} reads in mixed-canonical form 
at sites (orbitals) $\{l, l+1\}$ \cite{scho11}
\begin{equation}\label{PsiMPScan}
 \ket{\Psi} = \sum_{\boldsymbol{\sigma}} \sum_{a_1,\cdots,a_{L-1}}A_{1 a_1}^{\sigma_1}\cdots A_{a_{l-2} a_{l-1}}^{\sigma_{l-1}} M_{a_{l-1},a_{l+1}}^{\sigma_l \sigma_{l+1}} B_{a_{l+1} a_{l+2}}^{\sigma_{l+2}}\cdots B_{a_{L-1} 1}^{\sigma_L} \ket{\boldsymbol{\sigma}}\ ,
\end{equation}
where the MPS tensors $A^{\sigma_{l-1}}=\{A^{\sigma_{l-1}}_{a_{l-2}a_{l-1}}\}$\ and $B^{\sigma_{l-1}}=\{B^{\sigma_{l+2}}_{a_{l+1}a_{l+2}}\}$\ are left- and right-normalized, respectively. 
The action of the Hamiltonian $\hat{H}$ on the MPS state in mixed-canonical form can then be written as  
\begin{equation}\label{HPsiMPS}
\hat{H} \kPsi = \sum_{b_{l-1},b_l} \sum_{a'_{l-1},\sigma'_l,a'_l} L_{b_{l-1}}^{a_{l-1},a'_{l-1}} W_{b_{l-1},b_{l+1}}^{\sigma_l \sigma_{l+1},\sigma'_l \sigma'_{l+1}} R_{b_{l+1}}^{a_{l+1},a'_{l+1}} M_{a_{l-1},a_{l+1}}^{\sigma'_l,\sigma'_{l+1}} | a_{l-1} \rangle_A |\sigma_l \rangle |\sigma_{l+1} \rangle |a_{l+1} \rangle_B
\end{equation}
with the left and right basis states given by
\begin{equation}\label{PsiMPSA}
 | a_{l-1} \rangle_A = \sum_{\sigma_{1},\cdots,\sigma_{l-1}} \sum_{a_1,\cdots,a_{l-1}}A_{1 a_1}^{\sigma_1}\cdots A_{a_{l-2} a_{l-1}}^{\sigma_{l-1}} |\sigma_1,\cdots,\sigma_{l-1} \rangle
\end{equation}
and 
\begin{equation}\label{PsiMPSB}
 | a_{l+1} \rangle_B = \sum_{\sigma_{l+1},\cdots,\sigma_{L}} \sum_{a_{l+1},\cdots,a_{L}} B_{a_{l+1} a_{l+2}}^{\sigma_{l+2}}\cdots B_{a_{L-1} 1}^{\sigma_L} |\sigma_{l+1},\cdots,\sigma_{L} \rangle\ .
\end{equation} 
The left and right boundaries \cite{scho11,kell15a} are 
\begin{equation}\label{Lcontr}
 L_{b_{l-1}}^{a_{l-1},a'_{l-1}} = \sum_{\{a_i,b_i,a'_i;i<l-1\}} \left( \sum_{\sigma_1,\sigma'_1} A_{1,a_1}^{\sigma_1*} W_{1,b_1}^{\sigma_1,\sigma'_1}A_{1,a'_1}^{\sigma'_1} \right )\cdots \left( \sum_{\sigma_{l-1},\sigma'_{l-1}} A_{a_{l-2},a_{l-1}}^{\sigma_{l-1}*} W_{b_{l-2},b_{l-1}}^{\sigma_{l-1},\sigma'_{l-1}*} A_{a'_{l-2},a'_{l-1}}^{\sigma'_{l-1}} \right )
\end{equation}
and 
\begin{equation}\label{Rcontr}
 R_{b_{l+1}}^{a_{l+1},a'_{l+1}} = \sum_{\{a_i,b_i,a'_i;i>l+1\}} \left( \sum_{\sigma_{l+1},\sigma'_{l+1}} B_{a_{l+1},a_{l+2}}^{\sigma_{l+2}*} W_{b_{l+1},b_{l+2}}^{\sigma_{l+2},\sigma'_{l+2}} B_{a'_{l+1},a'_{l+2}}^{\sigma'_{l+2}} \right ) \cdots \left( \sum_{\sigma_{L},\sigma'_{L}} B_{a_{L-1},a_1}^{\sigma_{L}*} W_{b_{L-1},b_{1}}^{\sigma_{L},\sigma'_{L}*} B_{a'_{L-1},a'_{1}}^{\sigma'_{L}} \right )\ .
\end{equation}

In a variational optimization of $\ket{\Psi}$, we minimize the energy expectation 
value $\left<\Psi | \hat{H}  |\Psi\right>$ with respect to the entries of the MPS tensors under 
the constraint that the wave function is normalized, i.e., $\left<\Psi|\Psi\right> = 1$. Further assuming that the left and right 
boundaries were calculated from left- and right-normalized MPS tensors, this yields an eigenvalue equation \cite{ostl95,scho11}\ of the form
\begin{equation}\label{EigPsiMPS}
\boldsymbol{\mathcal{{H}}} v - \lambda v = 0 
\end{equation}
with the local Hamiltonian matrix $\boldsymbol{\mathcal{{H}}}$ at sites \{$l,l+1$\}\ after reshaping given by 
\begin{equation}\label{EigPsiMPSH}
H_{(\sigma_{l,l+1} a_{l-1}a_{l+1}),(\sigma'_{l,l+1} a'_{l-1}a'_{l+1})}  =  \sum_{b_{l-1},b_{l+1}} L_{b_{l-1}}^{a_{l-1},a'_{l-1}} W_{b_{l-1},b_{l+1}}^{\sigma_l \sigma_{l+1},\sigma'_l \sigma'_{l+1}} R_{b_{l+1}}^{a_{l+1},a'_{l+1}}
\end{equation}
and 
the vector $v$ collecting
\begin{equation}\label{EigPsiMPSV}
v_{\sigma_l \sigma_{l+1} a_{l-1} a_{l+1}} = M_{a_{l-1},a_{l+1}}^{\sigma_l \sigma_{l+1}}\ .
\end{equation}
Since we are often interested in only a few of the lowest eigenvalues $\lambda$, Eq.~(\ref{EigPsiMPS}) is best solved by an iterative eigensolver such as 
the Jacobi–Davidson procedure. For example, having obtained the lowest eigenvalue $\lambda_0$ and the corresponding eigenvector 
$v^0_{\sigma_l \sigma_{l+1} a_{l-1} a_{l+1}}$, the latter can be reshaped back to $M_{a_{l-1},a_{l+1}}^{\sigma_l \sigma_{l+1}}$\ which is then subject to a left- or right-normalization into $A^{\sigma_l}_{a_{l-1}a_{l}}$ or $B^{\sigma_{l+1}}_{a_{l}a_{l+1}}$ by a singular value decomposition in order to maintain the desired normalization structure. Given the optimized MPS tensors for sites $l$\ and $l+1$, the complete algorithm now \textit{sweeps}\ sequentially forth and back through the 'lattice' of sites consisting of the $L$\ spatial orbitals ordered in (arbitrary) form while optimizing each of the MPS tensors until convergence is reached. 

\subsubsection{Coupling Approaches}\label{sec:coupling-procedure}

The simultaneous optimization of the MPS parameters and orbital coefficients, which is a prerequisite for optimum convergence, can be accomplished in different ways \cite{wern80,wern85,know85} in our WMK optimization framework as will be discussed below. 

Following the idea of Werner and Meyer \cite{wern80}, we can write a coupled second-order energy expansion $E^{(2)}(\mT, v^{\Theta})$ setting out from Eq.~(\ref{E2origin}) in matrix form for a given point in the variational MPS optimization procedure of state $\Theta$\ as
\begin{align}\label{E2MPSorb}
E^{(2)}(\mT, v^{\Theta})  
            =  &\ \underbrace{v^{\Theta \dagger} \boldsymbol{\mathcal{{H}}} v^{\Theta}}_{E_{0}} + \frac{1}{2}{\rm tr}\left[\mT^\dagger \left(\bar{\textbf{A}}+\bar{\mB}\right)\right]+ 2 \sum_I \left(v^{\Theta}_I-\bar{v}^{\Theta}_I\right) {\rm tr} \left(\mT^\dagger \textbf{A}^I\right)\ ,
\end{align}
with $v$, $\boldsymbol{\mathcal{{H}}}$, \textbf{A}, and \mB\ defined as in Eqs.~(\ref{EigPsiMPSV}), (\ref{EigPsiMPSH}), (\ref{A}), and (\ref{Borig}), respectively. The bar in Eq.~(\ref{E2MPSorb}) indicates 
that the corresponding matrices are to be calculated with the one- and two-particle RDMs obtained for the initial MPS while the 
subscript $I$ denotes the $I$-th element of the coefficient vector $v^{\Theta}_I$ ($\bar{v}^{\Theta}_I$)\ as given in Eq.~(\ref{mps_ci})\ in Section \ref{sec:RDMder}. 
The matrix $\textbf{A}^I$ is defined \cite{wern80}\ as the corresponding matrix $\textbf{A}$ (cf.~Eq.~(\ref{A})) 
where the one- and two-particle RDM elements are to be replaced by the corresponding symmetrized RDM derivatives ${\gamma^I}_{ij}$\ (Eq.~(\ref{MPSderi1aT})) and ${\Gamma^I}_{ijkl}$ (Eq.~(\ref{MPSderi1bT})). Details on how to obtain the RDM derivatives for a given 
MPS state $\Theta$\ can be found in Section \ref{sec:RDMder}. 
By evaluating the first and second derivatives of the energy expression given by Eq.~(\ref{E2MPSorb}) 
with respect to the vector elements $v^{\Theta}_I$ and orbital rotation parameters $R_{ri}$ ($r > i$), we obtain the coupled NR equations 
\begin{equation}\label{MPSQCsiteC}
\left( \begin{array}{ccc} 
H_{R,R} &{H_{R,v}} \\
{H_{v,R}} & H_{v,v}  \\
\end{array} \right)
\left( \begin{array}{c} 
{{\mR}} \\ 
{{v^{\Theta}}} \\
 \end{array} \right)
\mbox{+}
\left( \begin{array}{c} 
g(\mR) \\ 
g(v^{\Theta}) \\
\end{array} 
\right)
\mbox{=}
\left( \begin{array}{c} 
0 \\ 
0 \\
\end{array}
\right)
\end{equation}
where the elements of the optimization parameters $\textbf{R}$ and $v^{\Theta}$ are both collected as vectors. The elements of the gradient $g$ and Hessian $H$ read as 
\begin{align}
g_{ri}\ =&\ \left( \frac{\partial E^{(2)}}{\partial R_{ri}}\right)_{\mR=0} = 2\left(\bar{A}_{ri}-\bar{A}^{\dagger}_{ri}\right)\ , \label{g_orb} \\
g_{v^{\Theta}_{I}} =&\  \left( \frac{\partial E^{(2)}}{\partial v^{\Theta}_{I}}\right)_{\bar{v}^{\Theta}=0} = \sum_{i,j} \langle i | h| j \rangle \gamma_{ij}^I + \frac{1}{2} \sum_{i,j,k,l}  (ij|kl)\Gamma_{ijkl}^I - E_{0}\ v_I\ , \label{g_mps}\\
H_{ri,sj} =&\ \left( \frac{\partial^2 E^{(2)}}{\partial R_{ri} \partial R_{sj}} \right)_{_{\mR=0}} 
= (1-\tau_{ri})(1-\tau_{sj}) \left(2\langle r|\bar{G}_{ij}|s\rangle - (\bar{A}_{rs}+\bar{A}^{\dagger}_{rs})\delta_{ij} \right)\ , \label{H_RR}\\
H_{v^{\Theta}_I,ri} =&\ \left( \frac{\partial^2 E^{(2)}}{\partial v^{\Theta}_{I} \partial R_{ri}} \right)_{{\bar{v}^{\Theta},\mR=0}} 
= 2(1-\tau_{ri}) (A^I_{ri}-\bar{v}^{\Theta}_I \bar{A}_{ri}) \label{H_cR} \ ,
\end{align}
and
\begin{align}
H_{v^{\Theta}_I,v^{\Theta}_J} =&\ \left(\frac{\partial^2 E^{(2)}}{\partial v^{\Theta}_{I} \partial v^{\Theta}_{J}} \right)_{_{\bar{v}^{\Theta},\mR=0}} 
= 2(H_{IJ}-E_{0}\delta_{IJ}) \ . \label{H_cc} 
\end{align}
where $H_{IJ}$\ is the $IJ$-th element of the (local) Hamiltonian matrix $\boldsymbol{\mathcal{{H}}}$\ defined in 
Eq.~(\ref{EigPsiMPSH}).
The Hessian given on the left-hand side of Eq.~(\ref{MPSQCsiteC}) corresponds to the second-order derivatives for a specific 
DMRG-SCF state $\Theta$\ but can be extended to the state-averaged case, which is particular useful for formulating a 
response or coupled-perturbed MCSCF-type approach \cite{helg86}. As outlined in Appendix B of Ref.~\citenum{wern80}, Eq.~(\ref{MPSQCsiteC})\ can then be solved in an iterative fashion. More details on the solution of Eq.~(\ref{MPSQCsiteC}) within our MPS/MPO framework and its relation to the calculation of analytical state-averaged DMRG-SCF gradients will be given in a forthcoming publication \cite{mayi17b}. 
For the remainder of this work, we chose to explore an alternative procedure which, in contrast to the NR-like approach above, 
exploits a direct-CI type step based on a second-order transformation of the Hamiltonian to obtain a relaxed set of MPS parameters. 
Moreover, as outlined in Ref.~\citenum{wern85}, the latter algorithm can quite easily be extended for an optimization of an energy average of an ensemble of electronic states, usually referred to as state-averaged optimization.

To this end, we express the second-order energy approximation $E^{(2)}(\mT)$\ for a given \mU\ as an expectation value \cite{wern87}
 \begin{equation}\label{E2mps}
 E^{(2)}(\mT) = \frac{\left<\Psi | \hat{H}^{(2)}  |\Psi\right>}{\left<\Psi|\Psi\right>}\ ,
\end{equation}  
where $\hat{H}^{(2)}$ is the second-order Hamiltonian defined as \cite{wern85,wern87}
 \begin{equation}\label{H2mps}
 \hat{H}^{(2)} = \sum_{i,j,\tau} \underbrace{\langle i |h^{(2)}| j\rangle}_{= h^{(2)}_{ij}}  \hat{c}_{i\tau}^{\dagger} \hat{c}_{j\tau}+ \frac{1}{2} \sum\limits_{\substack{i,j,k,l\\ \tau, \tau^{\prime}}}^{} (ij|kl)^{(2)}\ \hat{c}_{i\tau}^{\dagger} \hat{c}_{k\tau^{\prime}}^{\dagger} \hat{c}_{l\tau^{\prime}} \hat{c}_{j\tau}\ ,
\end{equation}  
with the second-order one-electron, 
 \begin{equation}\label{h2}
h^{(2)}_{ij} = (\mU^\dagger \textbf{h} \mU)_{ij}\ ,
\end{equation}  
and two-electron integrals,
 \begin{equation}\label{v2}
(ij|kl)^{(2)} = -(ij|kl) + (\mU^\dagger \textbf{J}^{kl}\mU)_{ij} + (\mU^\dagger \textbf{J}^{ij} \mU)_{kl} +(1+\tau_{ij})(1+\tau_{kl})(\mT^\dagger \textbf{K}^{ik} \mT)_{jl}\ .
\end{equation}
These integrals can be straightforwardly evaluated from the half-transformed integrals given in Eqs.~(\ref{eq:halfh})-(\ref{eq:halfK})\ by performing the second-half transformation for $\textbf{h}$, $\textbf{J}^{kl}$, and $\textbf{K}^{kl}$.
We then proceed with an iterative optimization (through one or several sweeps) 
of the MPS tensors as outlined in the previous Section \ref{sec:concepts}\ and based on the above second-order Hamiltonian $\hat{H}^{(2)}$\ expressed in MPO form (cf.~Eq.~(\ref{eq:MPO})). During a sweep, eigenvalue equations similar to Eq.~(\ref{EigPsiMPS})\ are solved
\begin{equation}\label{EigPsiMPS2}
\boldsymbol{\mathcal{{H}}}^{(2)} v^{(2)} - \lambda^{(2)} v^{(2)} = 0\ ,
\end{equation}
where the superscript $(2)$\ indicates that the MPS tensors are optimized with respect to the second-order Hamiltonian given in Eq.~(\ref{H2mps})\ in contrast to the 'standard' Hamiltonian (Eq.~(\ref{Hami_op}))\ employed in Eq.~(\ref{EigPsiMPS}). 
With the optimized $|\Psi^{(2)}\rangle$\ at hand, new one- and two-particle RDMs are calculated. Along with the half-transformed integrals (Eqs.~(\ref{eq:halfh})-(\ref{eq:halfK})), this allows us to calculate a new \mB\ matrix for the next micro iteration step\cite{wern85, know85, wern87}. 
Given a current set of parameters \mT\ and MPS tensors, convergence of the micro iterations is reached 
when both coupled Eqs.\ (\ref{UBBU}) and (\ref{EigPsiMPS2}) are satisfied simultaneously.  

The individual steps of our (coupled) second-order DMRG-SCF algorithm outlined above can be summarized as follows:
\begin{enumerate}
\item Set \textit{macro iteration} counter $t = 1$.
\item Given a set of MOs, carry out a 4-index transformation to obtain the one- and two-electron integrals in this MO basis. Perform a 
DMRG calculation within the active orbital space which yields the current MPS wave function.
\item With the current MPS and a given \mU\ (initial for $t = 1$, otherwise from the previous macro iteration), 
calculate the one- and two-particle RDMs $\boldsymbol{\gamma}$ and $\boldsymbol{\Gamma}$\ from which 
the operators $F^{ij}$\ and $G^{ij}$\ as well as the matrices $\textbf{A}$\ and $\textbf{B}$ can be obtained. 
Evaluate the approximate energy functional (cf.~ Eq.~(\ref{E2origin}))\ at the current expansion point and 
check for convergence with respect to $\delta E$. Set the \textit{micro iteration} counter $s = 1$.
\item Solve the linearized eigenvalue equation for variations of the non-redundant orbital rotations $\{\Delta R_{ri}\}$ by a step-restricted augmented (orbital) Hessian approach. Use $\{\Delta R_{ri}\}$ to update the rotation matrix \textbf{U}\ and calculate a new \textbf{B}\ matrix. Increase the micro iteration count, $s = s +1$. 
This step is repeated until either the stationarity condition in Eq.~(\ref{UBBU}) is satisfied within 
a given threshold or, alternatively, {$\mathbf{\Delta {\rm R}}$}\ becomes small; update \textbf{U}; if \textit{uncoupled} proceed to step 7 else 
continue with step 5.  
\item Use \textbf{U} and \textbf{T} to calculate the second-order one- and two-electron integrals (see Eqs.~(\ref{h2}) and (\ref{v2}))\ which define the corresponding second-order Hamiltonian $\hat{H}^{(2)}$. Based on $\hat{H}^{(2)}$ perform an update of the current MPS $\kPsi$ by carrying out additional DMRG sweep(s) to yield the second-order MPS $\ket{\Psi^{(2)}}$. With $\ket{\Psi^{(2)}}$ at hand, 
calculate new RDMs $\boldsymbol{\gamma}$ and $\boldsymbol{\Gamma}$.
\item Calculate \textbf{B} with the RDMs $\boldsymbol{\gamma}$ and $\boldsymbol{\Gamma}$\ obtained in the preceding step and 
the half-transformed integrals. If the stationary condition (Eq.~(\ref{UBBU})) is not satisfied, return to step 4. 
\item Update the MOs based on the final \textbf{U}, increase the \textit{macro iteration} counter by one, $t=t+1$, and continue with step 2. 
\end{enumerate}
The micro iterations comprise steps 4--6. In the \textit{uncoupled} case (denoted in the following as WMK), steps 5 and 6 are skipped as these  introduce the coupling between orbital rotation and MPS parameter optimization in the micro iterations (denoted as CP-WMK). 
The updated RDMs which enter the gradient evaluation in step 6 reflect the changes of the MPS subject to variations in the orbital rotation parameters.

In order to minimize a state-averaged energy $E_{0}$\ for an ensemble of $\theta$\ states within our 
WMK-based DMRG-SCF optimization scheme merely requires to replace the elements of the one- and two particle 
RDMs $\gamma_{ij}$ and $\Gamma_{ijkl}$ 
by their state-averaged counterparts, i.e., by
\begin{equation}\label{weighted_rdms1}
\gamma_{ij} = \sum_{\theta} w^\theta \gamma^\theta_{ij}
\end{equation}
and 
\begin{equation}\label{weighted_rdms2}
\Gamma_{ijkl} = \sum_{\theta} w^\theta \Gamma^\theta_{ijkl}\ ,
\end{equation}
where $w^\theta$ is a predefined weight for the $\theta$-th state. If needed, the sum $\sum\limits_\theta w^\theta$ is re-scaled to 1. 
Moreover, if the target states have different (spatial or spin) symmetries, they will be optimized independently in our DMRG program \qcm. 
Otherwise, each state is optimized in a sequential order by projecting out the previous lower and orthogonal states \cite{kell15a, knec16a} based on a Gram-Schmidt orthogonalization approach.

\subsection{RDM derivatives for an MPS wave function}\label{sec:RDMder}

In a configuration basis, elements of the one- and two-particle RDMs $\boldsymbol{\gamma}$\ and $\boldsymbol{\Gamma}$ 
(cf. Eqs.~(\ref{1eRDM}) and (\ref{2eRDM})) 
can be calculated from 
\begin{equation}\label{gamma}
\gamma_{ij} = \sum_{I,J} c_I \gamma_{ij}^{IJ} c_J
\end{equation}
and 
\begin{equation}\label{Gamma}
\Gamma_{ijkl} = \sum_{I,J} c_I \Gamma_{ijkl}^{IJ}c_J\ .
\end{equation}
Here, $c_I$ is the expansion coefficient for the $I$-th configuration $\Phi_I$ and the coupling coefficients $\gamma_{ij}^{IJ}$ and $\Gamma_{ijkl}^{IJ}$ 
are given by 
\begin{equation}\label{deri0}
\gamma_{ij}^{IJ} = \left< \Phi_I | \hat{c}_{i^\dagger} \hat{c}_{j} | \Phi_J \right>
\end{equation}
and
\begin{equation}\label{deri0b}
\Gamma_{ijkl}^{IJ} = \left< \Phi_I | \hat{c}_{i^\dagger} \hat{c}_{k^\dagger} \hat{c}_{l} \hat{c}_{j} | \Phi_J \right> .
\end{equation}
Starting from Eqs.~(\ref{gamma})\ and (\ref{Gamma}), the density matrix derivative with respect to $c_I$\ reads for the one-particle RDM 
\begin{equation}\label{deri1}
\tilde{\gamma}_{ij}^{I}(IJ) = \left(\frac{\partial \gamma_{ij}}{\partial c_I}\right)_{\{\textbf{c}\}} =  \sum_{I,J} \left[\frac{\partial {c_I} } {\partial c_I} \gamma_{ij}^{IJ} c_J + c_I \gamma_{ij}^{IJ} \frac{\partial {c_J} } {\partial c_I}\right] = \sum_{J}  c_J \gamma_{ij}^{IJ}
\end{equation}
and similarly for the derivative of the two-particle RDM 
\begin{equation}\label{deri1b}
\tilde{\Gamma}_{ijkl}^{I}(IJ) = \left(\frac{\partial \Gamma_{ijkl}}{\partial c_I}\right)_{\{\textbf{c}\}} = \sum_{J} \Gamma_{ijkl}^{IJ} c_J\ .
\end{equation}
Here, $(IJ)$ indicates that the derivative element has been obtained from a coupling coefficient corresponding to the matrix element $\left< \Phi_I | \hat{O} | \Phi_J \right>$. Interchanging $I$ and $J$ yields for the one-RDM derivative 
\begin{equation}\label{deri1_2} 
\tilde{\gamma}_{ij}^I (JI) = \sum_{J}  c_J \gamma_{ij}^{JI}\ ,
\end{equation}
such that we can define the final, symmetrized one-body RDM derivative $\gamma_{ij}^I$ as
\begin{equation}\label{deri1_12} 
\gamma_{ij}^I = \frac{1}{2} \left\{\tilde{\gamma}_{ij}^I (IJ) + \tilde{\gamma}_{ij}^I (JI) \right\}\ ,
\end{equation}
and equivalently for the symmetrized two-body RDM derivative $\Gamma^I$ 
\begin{equation}\label{MPSQCsite_deriRDM2}
\Gamma_{ijkl}^{I} =\frac{1}{2} \left\{\tilde{\Gamma}_{ijkl}^I (IJ) + \tilde{\Gamma}_{ijkl}^I (JI) \right\}\ .
\end{equation}

In order to find equivalent expressions for the RDM derivatives, Eqs.~(\ref{deri1_12}) and (\ref{MPSQCsite_deriRDM2}), in an MPS-based formalism, we write a given (converged) MPS state $\kPsi$\ in mixed-canonical form at any two sites $\{l, l+1\}$\ \cite{scho11} (see also Eq.~(\ref{PsiMPScan}))
\begin{align}\label{deri2}
\kPsi = &\ \sum_{\boldsymbol{\sigma}} \sum_{a_1,\cdots,a_{L-1}}A_{1 a_1}^{\sigma_1}\cdots A_{a_{l-2} a_{l-1}}^{\sigma_{l-1}} M_{a_{l-1},a_{l+1}}^{\sigma_l \sigma_{l+1}} B_{a_{l+1} a_{l+2}}^{\sigma_{l+2}}\cdots B_{a_{L-1} 1}^{\sigma_L} \left|\boldsymbol{\sigma} \right> \nonumber \\
= &\sum_{\sigma_l,\sigma_{l+1}}  \ v_{\sigma_l \sigma_{l+1} a_{l-1} a_{l+1}} \left| a_{l-1}\right>_A \left| \sigma_l \right> \left| \sigma_{l+1} \right> \left| a_{l+1}\right>_B\ ,
\end{align}
where the last equality follows from definitions given in Eqs.~(\ref{PsiMPSA}), (\ref{PsiMPSB}), and (\ref{EigPsiMPSV}). 
Note that the dimension of the coefficient vector $v_{\sigma_l \sigma_{l+1} a_{l-1} a_{l+1}}$ is $4\times 4 \times m_{a_{l-1}} \times m_{a_{l+1}}$\ and therefore at most $16m^2$. Hence, for computational efficiency, we will typically choose $l$\ and $l+1$\ to be the first two sites.  
Denoting the $I$-th element of $v_{\sigma_l \sigma_{l+1} a_{l-1} a_{l+1}}$ as $v_I$, and the $I$-th basis of $\left| a_{l-1}\right>_A \left| \sigma_l \right> \left| \sigma_{l+1} \right> \left| a_{l+1}\right>_B$ as $\ket{\Phi_I}$, we can write the MPS in a more compact form 
\begin{equation}\label{mps_ci}
\kPsi = \sum^{}_{I} v_{I} \ket{\Phi_I}\ .
\end{equation}
To arrive at the RDM derivatives, we then proceed as follows. The $I$-th element in $v_{\sigma_l \sigma_{l+1} a_{l-1} a_{l+1}}$\ is set to one,
while all other elements are set to zero, which yields a modified vector $\tilde{v}_{\sigma_l \sigma_{l+1} a_{l-1} a_{l+1}}$. 
The latter is reshaped back to $\tilde{M}_{a_{l-1},a_{l+1}}^{\sigma_l \sigma_{l+1}}$ (see Eq.~(\ref{EigPsiMPSV})) to give the modified MPS 
$\ket{\tilde{\Psi}^{I}}$. After that, the RDM derivatives can be obtained from 
\begin{equation}\label{MPSderi1a}
\gamma^{I}_{ij} = \frac{1}{2} \left\{ \left< \tilde{\Psi}^{I} | \hat{c}^{\dagger}_i \hat{c}_j |\Psi\right> +\left< \Psi| \hat{c}^{\dagger}_i \hat{c}_j |\tilde{\Psi}^{I} \right> \right\}
\end{equation}
and 
\begin{equation}\label{MPSderi1b}
\Gamma^{I}_{ijkl} =\frac{1}{2} \left\{ \left< \tilde{\Psi}^{I} | \hat{c}^{\dagger}_i \hat{c}^{\dagger}_j \hat{c}_k \hat{c}_l |\Psi \right> + \left< \Psi | \hat{c}^{\dagger}_i \hat{c}^{\dagger}_k \hat{c}_l \hat{c}_j |\tilde{\Psi}^{I} \right> \right\}\ ,
\end{equation}
or equivalently as 
\begin{equation}\label{MPSderi1aT}
\gamma^{I}_{ij} = \frac{1}{2} \left\{ \left< \tilde{\Psi}^{I} | \hat{c}^{\dagger}_i \hat{c}_j + \hat{c}^{\dagger}_j \hat{c}_i|\Psi \right> \right\}\ 
\end{equation}
and
\begin{equation}\label{MPSderi1bT}
\Gamma^{I}_{ijkl} =\frac{1}{4} \left\{ \left< \tilde{\Psi}^{I} | \hat{c}^{\dagger}_i \hat{c}^{\dagger}_k \hat{c}_l \hat{c}_j + \hat{c}^{\dagger}_k \hat{c}^{\dagger}_i \hat{c}_j \hat{c}_l + \hat{c}^{\dagger}_j \hat{c}^{\dagger}_l \hat{c}_k \hat{c}_i + \hat{c}^{\dagger}_l \hat{c}^{\dagger}_j \hat{c}_i \hat{c}_k |\Psi \right>  \right\} .
\end{equation}
From the form of Eqs.~(\ref{MPSderi1aT}) and (\ref{MPSderi1bT}) it becomes evident that the computational cost 
of an RDM derivative calculation can be up to $16m^2$ times a standard RDM evaluation depending on the actual dimension of the vector $v$ (\textit{vide supra}). In a coupled NR or AH algorithm, the derivatives need to be explictly re-calculated for an MPS wave-function \textit{ansatz} because the tensors $\{M^{\sigma_l}_{a_{l-1}a_l}\}$ (and therefore the vector $v$)\ are changing during a micro-iteration step. 
By contrast, in a configuration expansion one still needs to loop over all configurations but the coupling 
coefficients (see for example Eq.~(\ref{deri0})) can be stored which greatly simplifies an update of the 
corresponding RDM derivatives. 
As stated earlier, the numerical examples discussed in Section \ref{sec:numerical}\ 
focus on the performance of the (CP-)WMK optimization scheme which does not require the evaluation of RDM derivatives whereas we postpone a critical evaluation of the numerical performance of a (step-restricted) coupled AH algorithm for DMRG-SCF as well as 
analytic gradients for a state-average DMRG-SCF approach to a forthcoming publication \cite{mayi17b}.

\section{Numerical Examples}\label{sec:numerical}

\begin{figure}[!htp]
 \begin{center}
   \includegraphics[scale=1.00]{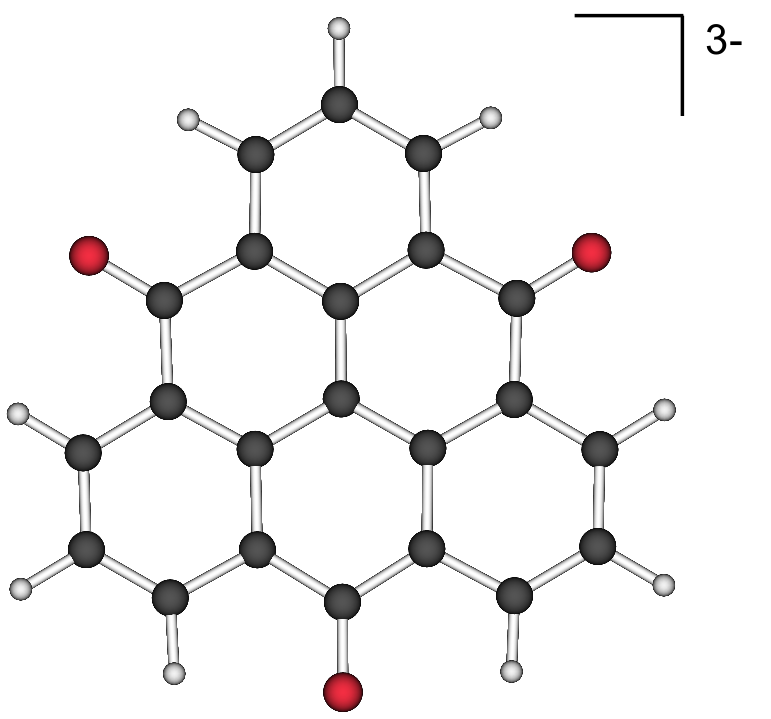}
 \end{center}
  \caption{Structure of the tri-anion of trioxytriangulene.}
\label{fig:mps-trioxytriangulene}
\end{figure}

We implemented the augmented Hessian and Werner-Meyer-Knowles optimization schemes as described in Section \ref{sec:method}. As default, the (in the limit of infinite $m$\ redundant) active-active orbital rotations\ 
are discarded in our implementation, but can be taken into account (denoted by an asterisk '$^{\ast}$', e.g., WMK$^{\ast}$) if requested. 
All DMRG calculations were carried out with our {\sc QCMaquis} DMRG software package \cite{kell15a,kell16,knec16a}
which is interfaced to a development version of the quantum chemistry software package {\sc Molcas} \cite{aqui15a}. 
We further took advantage of the latter interface to obtain reference DMRG-SCF data based 
on the Super-CI approach as implemented in {\sc Molcas}. 
Scalar-relativistic calculations for the chromiun dimer Cr$_2$ and copper dichloride CuCl$_2$ were carried out based on the second-order Douglas-Kroll-Hess (DKH2) Hamiltonian \cite{wolf02}, as implemented in the {\sc Molcas} program package \cite{reih04b}, 
in combination with ANO-RCC basis sets  \cite{widm90} and a triple-$\zeta$ contraction scheme (ANO-RCC-VTZP). 
For the [trioxytriangulene]$^{3-}$\ anion (see Fig.~\ref{fig:mps-trioxytriangulene}) we performed a ground state structure optimization imposing $D_{3h}$ symmetry 
with Gaussian09 \cite{g09} employing the long-range corrected hybrid $\omega$-B97XD density functional \cite{chai08} 
along with a correlation-consistent basis set of double-$\zeta$ quality (cc-pVDZ) \cite{dunn89}. 
The ensuing DMRG-SCF calculations were carried out with the ANO-S basis set and a double-$\zeta$\ contraction scheme (ANO-S-VDZP) \cite{pier95}. 
Starting from canonical Hartree-Fock orbitals, details concerning the choice of the active orbital space and the orbital ordering 
are given in the Supporting Information, in particular Tables S1-S3. Moreover, for simplicity all inactive orbitals are frozen in the DMRG-SCF optimization. Calculations were carried out on compute nodes equipped with two Xeon E5-2667 CPUs ($2\times 8$ cores @ 3.30 GHz) and 256 GB memory.

\subsection{Cr$_2$}\label{sec:cr2}

\begin{figure}[!htp]
\centering
   \includegraphics[scale=0.8]{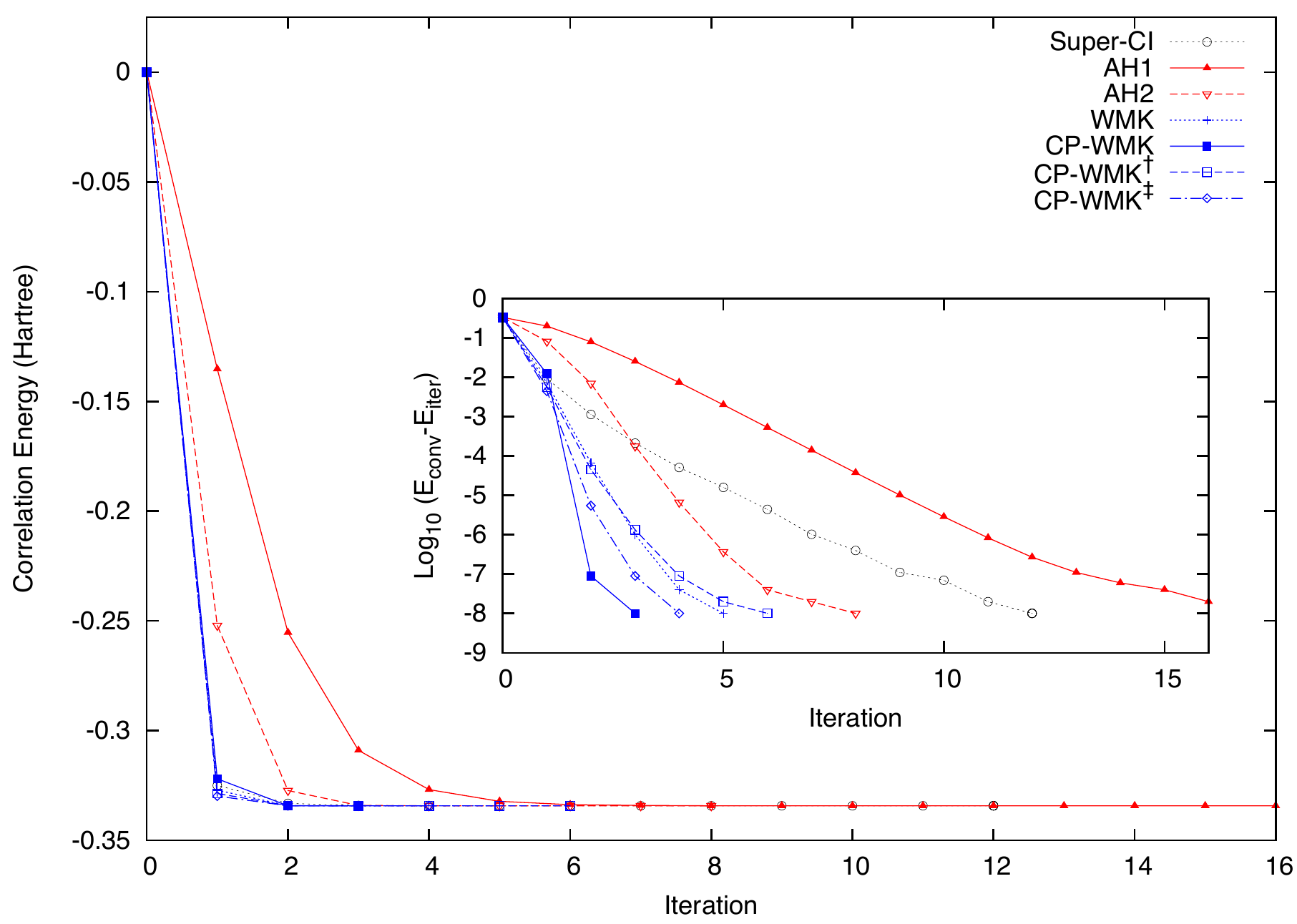}
  \caption{Convergence of DMRG(12,12)[500]-SCF for Cr$_2$ at a Cr-Cr internuclear distance $r_{\rm Cr-Cr} = 1.5$\ \AA\ 
  	using different orbital-optimization algorithms. The active space comprises the valence $3d4s$\ shells  
  	of each Cr atom. For details on the orbital-optimization algorithms see text.}
\label{Fig.Cr2_12e12o}
\end{figure}

Fig.~\ref{Fig.Cr2_12e12o} illustrates the convergence of our newly implemented DMRG-SCF approaches 
for the X$^1\Sigma_g^+$\ ground state of the chromium dimer (r$_{\rm Cr-Cr} = 1.5$ \AA). Absolute energies are compiled in Table S4 (see Supporting Information). The calculations were carried out with 
a 'minimal' active space for Cr$_2$\cite{roos94a,roos95a}, CAS(12,12), consisting of the 4$s$\ and 3$d$\ shell of each 
chromium atom (see Supporting Information for further details on the choice of active orbitals). 
By comparison to a conventional CI-driven CASSCF calculation, we confirmed that our DMRG-SCF 
calculations converged 
to the same solution for $m=500$. 
Given an energy convergence threshold of $\delta_{E}=10^{-8}$, both the Super-CI and standard 
AH algorithm (denoted as AH1 in Fig.~\ref{Fig.Cr2_12e12o}) require more than ten macro-iteration steps to reach 
the preset limit. In the latter case, the convergence criterion can be fulfilled after eight macro-iteration steps by introducing a 
step restriction (AH2; see Supporting Information for further details on the step-restriction algorithm).  
By contrast, resorting to the WMK optimization scheme, convergence is reached within five macro-iteration steps 
without MPS coupling (denoted as WMK) but further decreases to three macro-iteration steps by taking into account a 
simultaneous optimization of the MPS and orbital rotation parameters (CP-WMK) as described in Section \ref{sec:simulOPT}. 
In the remaining models dubbed as CP-WMK$^\dagger$ and CP-WMK$^\ddagger$ in Fig.~\ref{Fig.Cr2_12e12o}, 
we explored  the possibility of carrying out only a partial sweep rather than a full (or a few) sweep(s) for the MPS update 
(see step 5 in the WMK optimization algorithm sketched in Section \ref{sec:simulOPT})\ with the second-order 
Hamiltonian as given in Eq.~(\ref{H2mps}). 
In the case of CP-WMK$^\dagger$, only a single micro-iteration step (out of 11 for the full sweep) was performed 
whereas for CP-WMK$^\ddagger$\ the sweep procedure was terminated after three micro-iteration steps. 
As can be seen in Fig.~\ref{Fig.Cr2_12e12o} in comparison to the genuine CP-WMK model, one up to a few more macro iteration(s) are required for 
the  CP-WMK$^\dagger$\ and CP-WMK$^\ddagger$\ optimization approaches 
to reach convergence. In general, the computational overhead of additional macro iterations will outweigh the savings gained 
by an approximated MPS update in the micro iteration(s). Hence, in the remainder of this work we will only consider the CP-WMK model. 

\begin{figure}[!htb]
 \centering
   \includegraphics[scale=0.8]{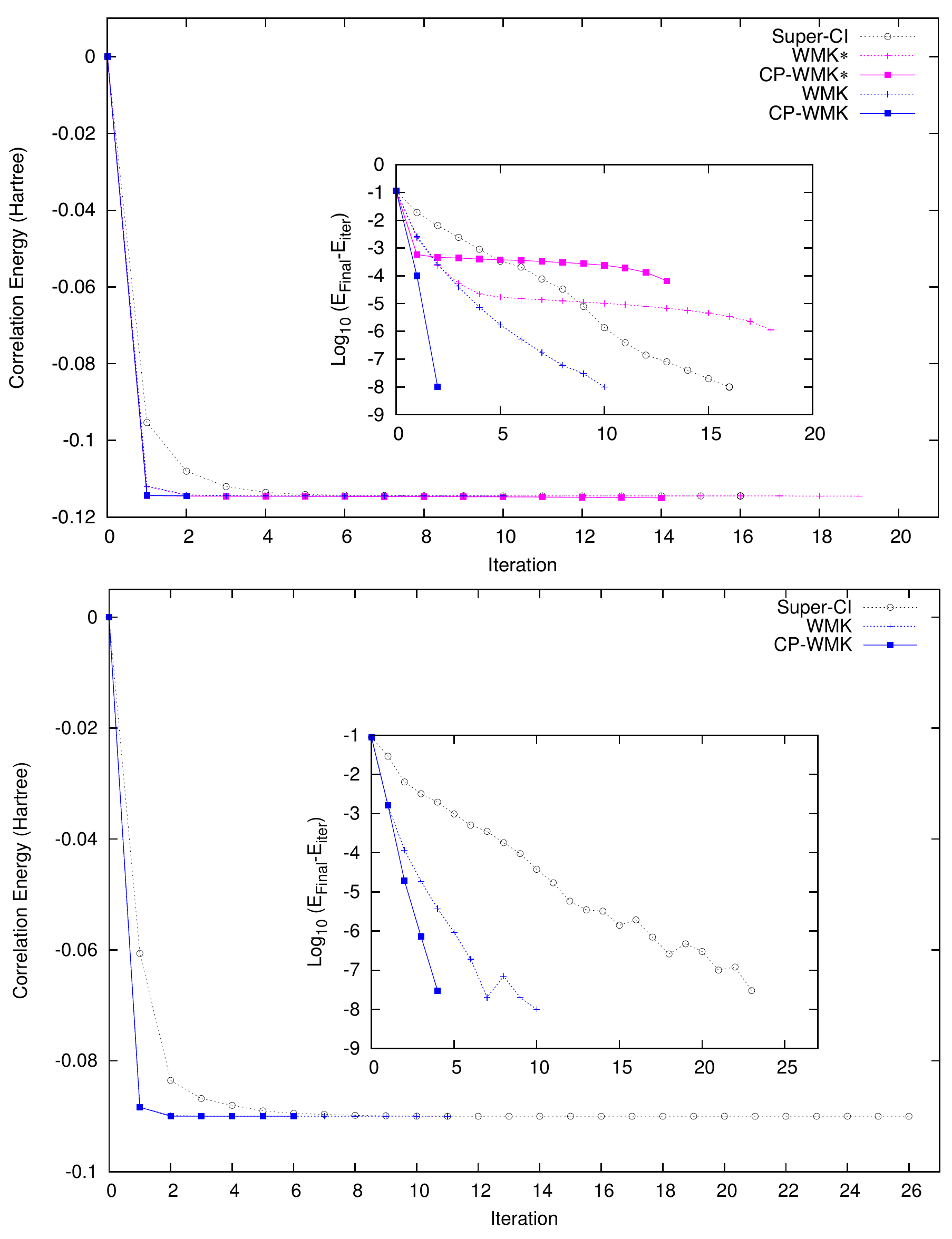}
 \vspace{-0.15cm}
  \caption{Convergence of DMRG(12,28)[1000]-SCF for Cr$_2$ at two representative Cr-Cr internuclear distances $r_{\rm Cr-Cr}$\ 
  	using different orbital-optimization algorithms. The active space comprises the valence $3d4s4p$\ shells  
  	and a double-$d$\ shell of each Cr atom. Top: $r_{\rm Cr-Cr}$ = 1.5 \AA. Bottom: $r_{\rm Cr-Cr}$ = 2.8 \AA. For details on the orbital-optimization algorithms see text.}
\label{Fig.Cr212e28oVE}
\end{figure}

Since it is well-known that the CAS(12,12) space is an inadequate starting point for 
subsequent multi-reference correlation approaches \cite{kura11,vanc16a}, we explored also the convergence properties of our DMRG-SCF approaches for the CAS(12,28) space proposed by Kurashige and Yanai \cite{kura11}\ at two different points along the potential energy curve, $r_{\rm Cr-Cr} = 1.5$ \AA\ and $r_{\rm Cr-Cr} = 2.8$ \AA. 
The CAS(12,28) active orbital space is a superposition of the CAS(12,12) space with the addition of the $4p$\ as well as the 
so-called double-$d$ shell for each Cr atom. 
In Ref.~\citenum{vanc16a}, the effect of correlating the semi-core $3s$ and $3p$ orbitals was studied recently by means of RASSCF/RASPT2 calculations and compared to recent DMRG-NEVPT2 data by Guo \textit{et al.} \cite{guos16}, who 
investigated the correlation effect of the $3s$ and 3p orbitals in the DMRG-SCF and subsequent NEVPT2 step. 
Summarizing both studies, it was concluded that the correlation effect of the $3s$ and $3p$ orbitals is negligible or at least smaller 
than the double-$d$-shell effect. Since we do not aim at a quantitative description of the Cr$_2$\ potential energy curve in this work, we focus in the following on the CAS(12,28) active space which comprises a double-$d$ shell for each Cr atom.
In view of this larger CAS, 
the number of renormalized states $m$\ was increased to $m = 1000$\ for the DMRG calculation. As shown in Fig.~\ref{Fig.Cr212e28oVE}, the CP-WMK approach allows again for fast convergence 
within six macro-iteration steps whereas the uncoupled optimization WMK scheme requires about ten macro-iteration steps. 
This is, however, still considerably less compared to the generally slow convergence of our Super-CI reference approach. 
The latter demonstrates that an account of higher-order terms in the orbital rotation parameters, as is the case for the WMK approaches,
ensures fast and reliable convergence of the multiconfigurational wave function. It becomes particularly beneficial at expansion points 
on the potential energy surface such as the stretched Cr-Cr configuration, where small orbital rotations between 
weakly occupied and secondary orbitals with little or no impact on the total energy can cause slow and oscillatory convergence. 

Up to this point we tacitly assumed that our DMRG optimization yields for a given $m$\ (initial guess for the MPS and orbital ordering) 
an MPS wave function of (near) full-CI quality which makes the wave function invariant with respect to inactive-inactive, 
active-active, and secondary-secondary orbital rotations as well as to the ordering of the active orbitals. Consequently, active-active 
rotations are not considered in the CP-WMK optimization scheme. 
For larger active orbital spaces convergence to the true full-CI solution requires, however, sufficiently large values of $m$\ in the DMRG calculation. As shown in Fig.~\ref{Fig.Cr212e28oVE}, explicitly taking into account active-active orbital rotations 
in the uncoupled WMK$^{\ast}$\ and coupled CP-WMK$^{\ast}$ optimization schemes breaks the convergence efficiency previously 
observed for (CP-)WMK. The final convergence criterion ($\delta_{E}=10^{-8}$) could not be reached even after ten macro-iteration steps. 
While being redundant (and therefore zero by definition) for FCI wave functions, the occurrence of (small) active-active 
rotations clearly indicates that for the given CAS(12,28)\ our $m$\ value is not sufficient to converge the DMRG wave function to (near)-full CI quality. This is also reflected in the final total energies which are compiled in Table S5 (see Supporting Information) for different 
optimization approaches with a fixed $m=1000$\ and in Table S6 for the Super-CI and CP-WMK approach at $r_{\rm Cr-Cr} = 2.8$ \AA\ with 
$m$\ values ranging from 500 to 3000. 

Focusing on a comparison of CP-WMK and CP-WMK$^{\ast}$, we note that the inclusion of active-active rotations results 
in an energy lowering in the sub-mH range. As pointed out by Zgid and Nooijen \cite{zgid08}, this could be further exploited to 
explicitly minimize the energy with respect to orbital rotations in the active space but requires tailored algorithms as 
the slow convergence of our unmodified CP-WMK$^{\ast}$\ approach clearly indicates. 
Consequently, a better and more sustainable solution will be to discard the active-active orbital rotations and, if possible, 
increase the number of renormalized states to larger values both of which will allow for fast and reliable convergence. 
The latter has not only the potential to converge in a unbiased fashion to a (stationary) point on the parameter surface close 
to the true (local) minimum \cite{zgid08}\ but can also provide a sufficiently accurate starting point 
for post-DMRG-SCF approaches. A closer look at Table S6 (see Supporting Information) reveals for increasing $m$\ 
(i) a smooth convergence of the DMRG-SCF energies with the $m = 3000$\ result likely to be in reach of the FCI reference 
within $\mu$H accuracy, and (ii) a consistent agreement of the Super-CI and CP-WMK results within sub-$\mu$H accuracy, all of which corroborates the above conclusions.

\subsection{CuCl$_2$}\label{sec:cucl2}

All of the chromium dimer results discussed in the previous section were obtained for state-specific DMRG-SCF calculations. 
However, as it is, for example, the case for photo- and transition metal chemistry \cite{roca12}, it is often useful in a multiconfigurational approach 
to optimize the energy average for all states under consideration rather than each state individually. 
Although the optimized orbitals will then constitute a compromise, a state-averaged \textit{ansatz} can for example 
help to prevent root-flipping along the reaction coordinate in a photochemical process that involves more than one electronically excited state \cite{harv14}. 
We therefore studied the performance of our second-order CP-WMK implementation for state-specific and state-averaged DMRG-SCF calculations of different low-lying electronic states of the linear, centrosymmetric CuCl$_2$ molecule at a stretched 
Cu-Cl internuclear distance of 2.154 \AA. The number of renormalized states $m$\ was set to 500 in all calculations. 
 \begin{table}[!hbp]
 \centering 
 \caption{\label{tab-CuCl2} Comparison of the convergence of the CP-WMK approach for state-specific DMRG(21,17)[500]-CASSCF calculations of various electronic states of the linear, centrosymmetric CuCl$_2$ molecule (r$_{\rm Cu-Cl}=2.154$ \AA). All values are in $E_{h}$.
 }
  \begin{tabular}{cccccccc}
  \hline
  \hline   
       &&   \multicolumn{6}{c}{Energy difference} \\
   \cline{3-8}
Iter.  &&  $^2\Sigma_g^+$ & $^2\Pi_g$ & $^2\Sigma_u^+$ & $^4\Sigma_g^+$ & $^4\Pi_g$ & $^4\Delta_g$ \\ 
  \cline{1-8}
   1    && -0.24737269 & -0.22901085 & -0.23391455 & -0.36005375 & -0.30273352 & -0.29342618  \\
   2    && -0.00049572 & -0.00095797 & -0.00093486 & -0.00155592 & -0.00148766 & -0.00269265    \\
   3    && -0.00000000 & -0.00000000 & -0.00000000 & -0.00000000 & -0.00000000 & -0.00000831   \\
  \hline
  \hline
  \end{tabular}
\end{table}

Table~\ref{tab-CuCl2} compiles the state-specific data, whereas state-averaged results are listed in Table \ref{tab-CuCl2_SA}. 
'\textsc{state-averaged-1}' denotes a DMRG-SCF optimization 
targeting with equal weights the lowest four $^2\Sigma_g^+$ states and one $^2\Delta_g$\ state within the 
a$_g$\ point group irrep of D$_{2h}$. 
The second set (\textsc{state-averaged-2})\ denotes a simultaneous DMRG-SCF optimization of the lowest two $^2\Sigma_u^+$\ and 
$^2\Pi_u$\ charge-transfer states \cite{zouw09}\ with equal weights. In addition, three low-lying, so-called ligand-field states \cite{zouw09}, namely 
$^2\Sigma_g^+$, $^2\Pi_g$\ and $^2\Delta_g$\ are included with a small weight of 0.1. This composition was shown to avoid artificial symmetry-breaking in structure optimizations of the charge-transfer states \cite{zouw08, zouw09}. 
The active space comprises 21 electrons in 17 orbitals (Cu 3$d$4$s$3$d^{\prime}$ and two sets of Cl 3$p$\ orbitals, see also Section I.B in the Supporting Information). A double-$d$\ shell consisting of a linear combination of Cu 3$d$ and 4$d$ orbitals and 
denoted as 3$d^{\prime}$ was added to the Cu valence orbital space. The latter are known to be important for a balanced description of the differential electron correlation effects in the 3$d^{10}$ and 3$d^9$ super-configurations of Cu in ground- and excited states \cite{fisc77}. 

Summarizing Tables \ref{tab-CuCl2} and \ref{tab-CuCl2_SA}, we note a fast and reliable convergence of our DMRG-SCF calculations 
in either case, state-specific and state-averaged, within two to three macro-iteration steps based on the second-order CP-WMK 
optimization scheme. 
In addition, we observe already in the first macro iteration a rapid decrease of the rotation angles between active and secondary 
orbitals resulting in $\{\Delta R_{ri}\}$\ values less than 0.1 after approximately five micro-iteration steps. 
This underlines the significance of the double-$d$\ shell 3$d^{\prime}$ orbitals for a reliable description of the 
super-configurations of Cu\ in the ground- and excited states of CuCl$_2$. Compared to the RASSCF/RASPT2 data by 
Zou and Boggs \cite{zouw09}, all our DMRG-SCF calculations predict the $^2\Sigma_g^+$\ rather than the $^2\Pi_g$\ state as electronic ground state with a vertical energy gap of approximately 1200 cm$^{-1}$. We ascribe this apparent discrepancy primarily 
to dynamical electron correlation effects which are, despite the sizable active orbital space, only partially accounted for 
in our multiconfigurational calculations. 
 \begin{table}[!hbt]\centering 
 \caption{\label{tab-CuCl2_SA} Comparison of the convergence of the CP-WMK approach for state-averaged DMRG(21,17)[500]-CASSCF calculations averaging over electronic states of the same spatial point group irrep (\textsc{state-averaged-1}) and over electronic states belonging to different point group irreps (\textsc{state-averaged-2}) in the linear, centrosymmetric CuCl$_2$ molecule (r$_{\rm Cu-Cl}=2.154$ \AA). See text for details on the composition of the state-averaged state selection. All values are in $E_{h}$.
 }
  \begin{tabular}{ccccccc}
  \hline
  \hline   
       &&   \multicolumn{2}{c}{\textsc{state-averaged-1}}  && \multicolumn{2}{c}{\textsc{state-averaged-2}} \\
   \cline{3-4} \cline{6-7}
Iter.  &&  energy difference &  step lengths$^b$  &&   energy difference &  step lengths$^b$ \\ 
  \cline{1-7}
   1    && -0.25505955 & 2.7271  && -0.22833872 & 2.7171 \\
   2    && -0.00073400 & 0.1319  && -0.00084603 & 0.1587 \\
   3    &&  0.00000000 & 0.0002  &&  0.00000000 & 0.0000 \\   
  \hline
  \hline
  \multicolumn{7}{l}{$^a M$=500 for each state in the state-averaged space; states are solved consecutively.}\\
  \multicolumn{7}{l}{$^b$ Step lengths are defined as $(\sum\limits_{n>m} T_{nm}^2)^{1/2}$.}
  \end{tabular}
\end{table}

\subsection{Trioxytriangulene}\label{sec:trioxy} 

The tri-anion of trioxytriangulene, shown in Fig.~\ref{fig:mps-trioxytriangulene}\ and denoted in the following as \trio, belongs to the class of non-Kekul{\'e} polybenzenoid aromatic compounds \cite{rand03}. A characteristic feature of these unique $\pi$-biradicals is their triplet ground state \cite{kurr93}. Despite having an even number of carbon (ring)-atoms and an even number of $\pi$-electrons, 
two of the $\pi$-electrons cannot pair due to the molecule's geometry such that no classical Kekul{\'e} structure can be written \cite{clar62}. 
Hence, they are of potential interest as organic magnets. 
In contrast to the most prominent member of the family of non-Kekul{\'e} polybenzenoid aromatic compounds, namely triangulene, 
also known as Clar's hydrocarbon, which has been, until very recently \cite{pavl17}, experimentally elusive since its prediction \cite{clar53},   
\trio\ was first synthesized successfully in 1993 \cite{alli93}. 
Its electronic triplet ground-state character is supported by both electron paramagnetic resonance measurements \cite{alli93} 
and molecular-mechanics valence bond calculations based on a Heisenberg model Hamiltonian \cite{bear94a}. The latter predicted a 
triplet stabilization by approximately 20 kcal/mol compared to the open-shell singlet configuration. 
As Fig.~\ref{fig:mps-trioxytriangulene}\ suggests, \trio\ has a spatially extended valence $\pi$-conjugation system that includes contributions from all carbon as well as oxygen atoms. Hence, our active orbital space for \trio\ comprises all valence $\pi/\pi^\ast$\ orbitals resulting in a CAS(28,25)\ active orbital space. 

\begin{figure}[tbh]
 \centering
   \includegraphics[scale=0.8]{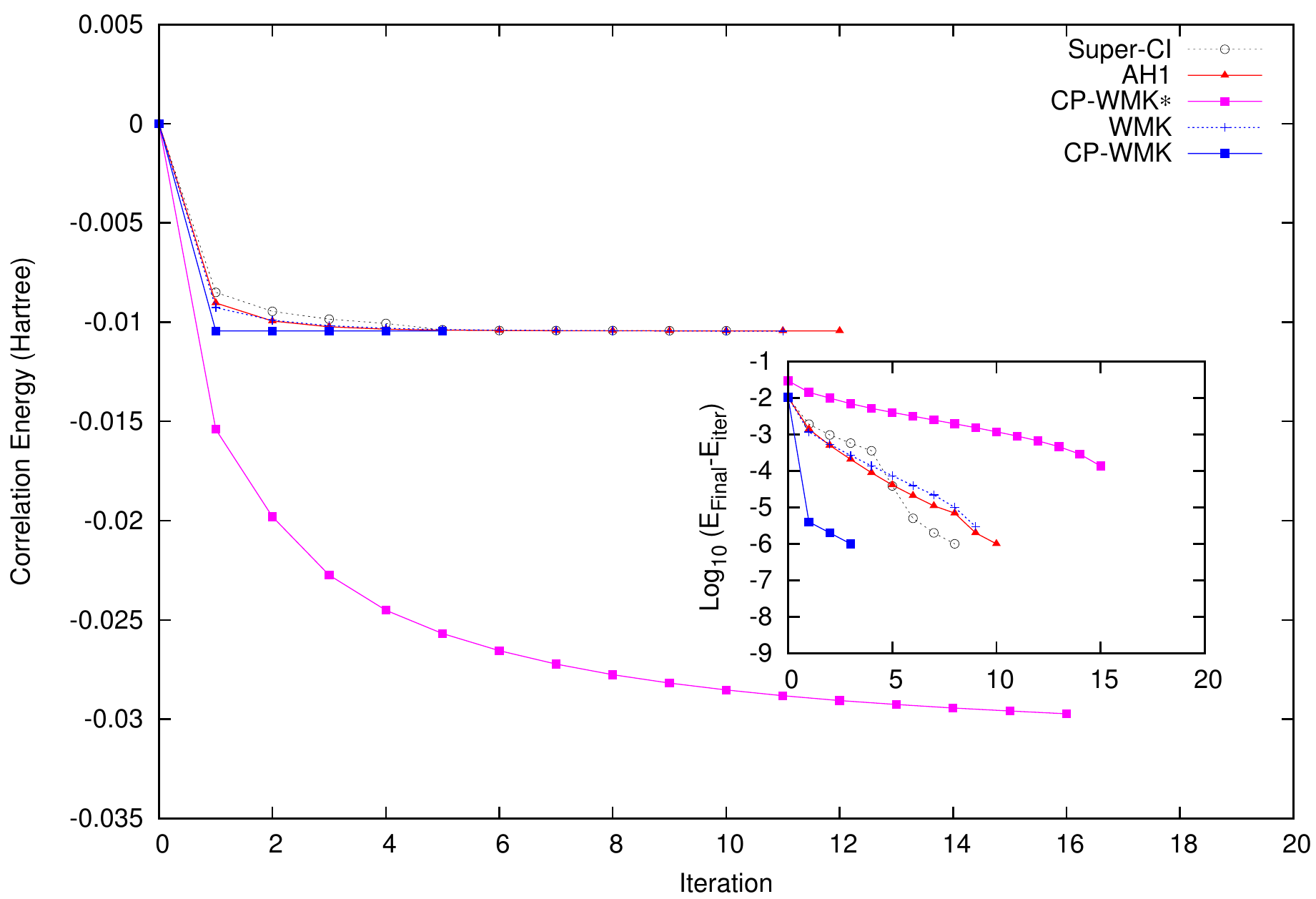}
  \caption{Convergence of DMRG(28,25)[1000]-SCF calculations of the $^3$B$_1$\ state of \trio. 
  	The convergence threshold for the energy is $\delta E = 10^{-6}$ Hartree.}
\label{Fig.trianion}
\end{figure}

Fig.~\ref{Fig.trianion}\ summarizes the DMRG-SCF convergence of different optimization approaches for the triplet $^3$B$_1$ ground state \footnote{We adopt here the notation for C$_{2v}$\ symmetry which has been employed for the calculations instead of the full symmetry C$_{3h}$\ which is neither available in \texttt{Molcas}\ nor in our program package.} of \trio. 
As can be seen from Fig.~\ref{Fig.trianion}, our second-order CP-WMK DMRG-SCF algorithm (blue filled squares in Figure \ref{Fig.trianion}) 
allows for convergence of the energy within four macro-iteration steps for $\delta E = 10^{-6}$\ which requires 
half the number of macro iterations compared to the Super-CI (grey open circles) and uncoupled WMK (light blue ``+" symbols) approaches 
with respect to the same convergence criteria. 
In Ref.~\citenum{kell15a}, it was concluded for a topological similar snippet of graphene (denoted as ``G2") 
that, in order to reach (near) FCI quality for the CAS wave function, requires for non-localized orbitals 
(which is generally, but not necessarily has to be \cite{mcdo86} the case in a DMRG-SCF optimization) $m$\ values of at least $m = 3000$ 
and possibly higher. This slow convergence is a consequence of the 2D-like, spatially extended correlation topology of 
the conjugated $\pi/\pi^\ast$\ system in G2 which is similar to the present situation in \trio. 
Moreover, in a study of arenes of different ``width" \cite{oliv15a}, i.e., with increasing number of condensed tetracenes, 
similar conclusions were drawn with respect to energy convergence and orbital-basis locality for given values of $m$. 
It might therefore be worthwhile to apply to such systems the recently presented quantum-chemical version of the so-called 
tree-tensor network states approach which would allow to explicitly consider the (multi-orbital) entanglement topology 
for the wave function optimization \cite{barc11a,naka13,szal15,murg15}. 
Since the maximum $m$\ value was set to $m=1000$\ in our DMRG-SCF calculations, 
an explicit account of active-active orbital rotations marked by an asterisk ``$^{\ast}$" in Fig.~\ref{Fig.trianion}\ was expected to have an impact not only on the MPS wave function optimization but also on the orbital optimization since both are coupled. 
Inspection of Fig.~\ref{Fig.trianion} unequivocally reveals, that a DMRG-SCF optimization based on the CP-WMK$^{\ast}$\ approach 
leads to a substantial energy lowering and slower convergence radius with respect to CP-WMK which does not account for active-active rotations. 
Hence, the presence of (by definition) redundant active-active orbital rotations in a DMRG-SCF optimization 
based on a CAS orbital space model could be exploited as a criterion to assess the quality of an MPS 
wave function obtained with a given $m$\ value. Such an estimate could be of particular value for 
the assessment of subsequent multi-reference perturbation calculations based on, for example, DMRG-NEVPT2 \cite{guos16,roem16,soko16,frei16}. In the latter case, invariance of the CAS reference wave function with respect to active-active rotations is assumed\ and its violation could give rise to numerical 
instabilities in the perturbation approach \cite{frei16}.

In order to investigate the triplet ground-state nature of \trio, we additionally performed a DMRG-SCF optimization for the lowest singlet $^1$A$_1$\ state. From the latter calculation, we estimate a 
triplet-singlet ($^3$B$_1$-$^1$A$_1$) gap of 10.3 kcal/mol which --- albeit being approximately 10 kcal/mol lower than the previous estimate by Bearpark \textit{et al.}\ \cite{bear94a} for \trio\ as well as for the triplet/singlet splitting reported for triangulene in Ref.~\citenum{dasa16a} --- 
confirms the expected triplet ground-state character of \trio.

\subsection{Computational costs}\label{sec:numerical-efficiency}
\begin{figure}[!htb]
\centering
\includegraphics[scale=0.7]{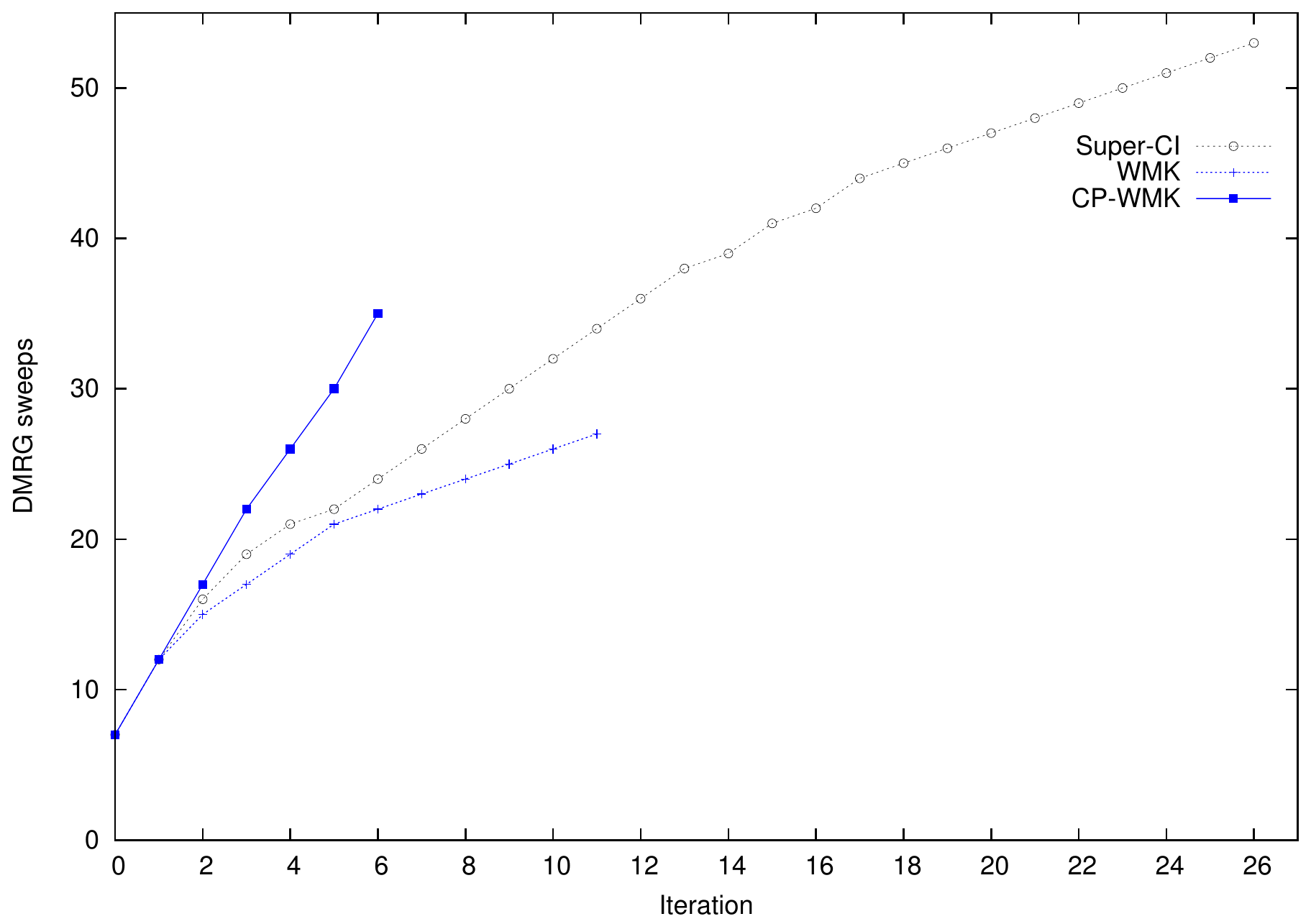}
\caption{Accumulated number of DMRG sweeps as a function of macro iterations for DMRG(12,28)[1000]-SCF calculations of the X$^1\Sigma_g^+$\ state of the Cr$_2$\ diatom carried out with different optimization algorithms. For further computational details, see Section \ref{sec:cr2}.}
\label{Fig.macro-sweeps}
\end{figure}

In the present work, we employ, as stated above, for simplicity a frozen-core approximation for all inactive orbitals, for example, 
all elements for inactive-active and inactive-secondary rotations are zero by construction. In doing so, the computational cost of the 
four-index integral transformation is significantly reduced since no two-electron integrals with inactive indices need to be evaluated. 
Clearly, this approximation does not allow us to obtain DMRG-SCF results of quantitative accuracy and we will assess its effect in 
our forthcoming publication \cite{mayi17b}, in particular since lifting the frozen-core approximation will be essential for an accurate 
analytic evaluation of (excited-state) state-average DMRG-SCF gradients. 
That said, we focus in the following on a comparison of the second-order WMK and CP-WMK approaches with the first-order Super-CI approach in terms of the number of DMRG sweeps required to converge the DMRG-SCF wave function. 
Within the frozen-core approximation, the computational cost of a macro iteration and the enclosed micro iterations is dominated 
by the active space solver which scales with DMRG as active space solver as $\mathcal{O}(m^3L^3) + \mathcal{O}(m^2L^4)$\ for a 
DMRG sweep. In Fig.~\ref{Fig.macro-sweeps}, we show the 
accumulated number of DMRG sweeps as a function of macro iterations required to converge the X$^1\Sigma_g^+$\ state of Cr$_2$\ 
by means of DMRG(12,28)[1000]-SCF calculations at an Cr-Cr internuclear distance of $r_{\rm Cr-Cr}=2.8$ \AA. 
As can be seen from Fig.~\ref{Fig.macro-sweeps}, the number of macro iterations and, more importantly, the number of DMRG-sweeps 
in the Super-CI approach exceeds by far those encountered for the (CP-)WMK approaches. 
This is also reflected in the total wall time which ranges from 7600 s (WMK) to 10500 s (CP-WMK) up to 29000 s for the two-step 
approach. Moreover, in Table \ref{tab.macro-micro-sweeps}\ we compare the convergence performance of our second-order 
WMK and CP-WMK approaches with the first-order Super-CI approach for the $^3$B$_1$\ state of \trio\ and the X$^1\Sigma_g^+$\ 
state of Cr$_2$. Considering first the Cr$_2$\ example, CP-WMK requires in total 73 micro iterations of which 13 are coupled micro iterations that require additional DMRG sweeps to optimize the MPS wave function for the approximate second-order Hamiltonian $\hat{H}^{(2)}$\ (cf.~Eq.~(\ref{H2mps})) as well as the evaluation of one- and two-particle RDMs. By contrast, the uncoupled WMK approach achieves convergence with less micro iterations (52) in combination with fewer DMRG sweeps than the CP-WMK approach which explains the lower total wall time. 
However, turning next to the optimization of the ground state of \trio, we find that, although the total number of micro iterations (54) for CP-WMK 
is almost twice the number of micro iterations for WMK (29), the total number of DMRG sweeps --- which dominates the overall computational cost --- 
is comparable. Interestingly, in this particular case the Super-CI approach 
leads to convergence of the DMRG-SCF wave function within 11 macro iterations comprising 46 DMRG sweeps to be compared with 12 macro iterations and 45 DMRG sweeps for the uncoupled WMK approach. The latter illustrates that the convergence rate of the first- and 
(uncoupled) second-order DMRG-SCF approaches may vary notably with the problem under consideration. Only by taking advantage of a coupled second-order a fast convergence within a few macro iterations can be ensured. This will be particular valuable for the study of 
extended molecular systems with a large number of atom orbital (AO) basis functions without the frozen-core approximation 
where the AO-to-MO four-index integral transformation required for each macro iteration will become a second bottleneck. 

\begin{table}[!htb]
\caption{Number of macro iterations, micro iterations and number of DMRG sweeps to converge the $^3$B$_1$\ state of \trio\ and the X$^1\Sigma_g^+$\ state of Cr$_2$ at an Cr-Cr internuclear distance of $r_{\rm Cr-Cr}=2.8$ \AA. 
For the CP-WMK algorithm, the number of micro iterations with coupled orbital and MPS optimization (steps 5 and 6 in the optimization algorithm outlined in Section \ref{sec:coupling-procedure})\ is given in parenthesis.}
\label{tab.macro-micro-sweeps}
\centering
\begin{tabular}{lccccccccc}\hline\hline
&&\multicolumn{3}{c}{\trio/CAS(25,28)} && \multicolumn{3}{c}{Cr$_2$/CAS(12,28)}\\ \cline{3-5} \cline{7-9}
&& Macro & Micro & DMRG sweeps && Macro & Micro & DMRG sweeps \\ \hline
Super-CI && 11 & 84$^a$ & 46 && 26 & 788$^a$ & 53 \\
WMK       && 12 & 29 & 45 && 11 & 52 & 27 \\
CP-WMK && 6 & 54 (19) & 44 && 6 & 73 (13) & 35 \\ \hline
\multicolumn{9}{l}{$^a$Number of SX iterations in the Super-CI algorithm of \texttt{Molcas}.}
\end{tabular}
\end{table}
\section{Conclusions and Outlook}\label{sec:conclusions}

	In this work, we presented a second-order coupled DMRG-SCF algorithm based on the Werner-Meyer-Knowles (WMK) optimization 
	scheme. Our quadratically convergent DMRG-SCF approach complements existing implementations for traditional CI-type wave 
	functions while generalizing earlier DMRG-SCF approaches to a fully second-order optimization framework. We illustrate two different possibilities to formulate a coupled MPS and orbital optimization scheme which in one case requires the calculation of derivatives of the one- and two-particle RDMs. By comparison to a CI-type wave function approach, we present a basic scheme for the evaluation of RDM derivatives in an MPS wave function framework.
	
	Quadratic convergence is achieved in our DMRG-SCF approach through extended micro iterations in which additional DMRG optimizations are carried out based on a second-order approximated Hamiltonian. Consequently, the MPS wave function is allowed to relax with respect to the orbital rotations predicted in the preceding micro iteration. Hence, by formulating a feedback loop between MPS wave function and 
	orbital rotation parameter updates, we iteratively solve the nonlinear equations defining the stationary conditions for the DMRG-SCF 
	optimization procedure simultaneously in a given macro-iteration step. We observed that in our DMRG-SCF implementation 
	of the WMK optimization scheme only a few MPS updates are required within the micro iterations in order 
	to significantly improve the convergence radius. The efficiency reported for the original formulation of the WMK optimization scheme 
	for CI-type wave functions carries over to our MPS-based optimization framework. In all examples studied in this work, three to six macro-iteration steps were sufficient to reach the final solution which is by a factor three to six less than the number macro-iteration 
steps needed by a second-order (orbital only) Augmented-Hessian or first-order Super-CI optimization approach. 
A fast and stable convergence within few macro iterations will be particularly 
	beneficial for addressing a multiconfigurational orbital optimization problems which not only necessitate large active orbital 
	spaces but also employ extended atomic orbital basis sets. Each macro-iteration step requires a four-index transformation of two-electron 
	integrals (with two general indices) which could easily become a computational bottleneck in those cases. 
	We will therefore tackle this issue explicitly in a future work by formulating a Cholesky-decomposition driven DMRG-SCF 
	implementation of the WMK optimization algorithm outlined in the present work. 
	
	The possibility to optimize a target wave function in a state-specific or state-averaged approach adds 
	an essential flexibility to our DMRG-SCF implementation which will open up for a manifold of applications in different contexts, 
	for example in photochemistry and transition-metal chemistry. 
	We demonstrated in this work the applicability of our second-order DMRG-SCF implementation for two transition-metal compounds, 
	namely Cr$_2$\ and CuCl$_2$, as well as for the tri-anion of trioxytriangulene, a prototypical non-Kekul{\'e} polynuclear aromatic 
	compound with a triplet electronic ground state. 
	By discarding active-active orbital rotations which -- strictly speaking -- are redundant in a complete-active space model only 
	for DMRG wave function of full-CI quality, we were able to converge ground and excited states in all considered examples within 
	at most four macro-iteration steps. Moreover, various state-averaged models either within a given (spatial or spin) symmetry or 
	across symmetries become feasible as illustrated by state-averaged DMRG-SCF calculations for low-lying excited states 
	of CuCl$_2$\ exhibiting different Cu 3$d$\ occupation patterns. 
	
	Striving ultimately for photochemical applications, the calculation of excited-state gradients for state-averaged wave functions 
	requires in a linear-response formalism analytic energy derivatives for a DMRG-SCF wave function similar to those presented 
	in this work. Work along this direction is currently in progress in our laboratory. 

At the time of submission of this work, we became aware of a paper on a topic similar to this work 
which was first uploaded to the arXiv preprint repository \cite{sunq16b}\ and later substantially revised \cite{sunq17b}. 
In this AO-driven implementation several approximations to the solution of Eq.~(\ref{MPSQCsiteC}) 
(cf.\ Eq.\ (16) in Ref.~\citenum{sunq17b}) were discussed and the so-called DEP1 approximation was considered 
to yield sufficiently fast and reliable convergence (for large-scale cases). In contrast to DEP1, which takes into account only a first order \mT\ expansion for the orbital gradient and CI Hamiltonian, we advocate in this work to take advantage of the coupled WMK approach (in an MO-basis formulation) which corresponds to a second-order expansion in \mT\ for the orbital gradient and CI Hamiltonian. We note that s similar approach dubbed as DEP2+ was also investigated 
for smaller active spaces and a full CI approach as active space solver in Refs.~\citenum{sunq16b,sunq17b}\ but not explored for larger active spaces and in combination with DMRG as active space solver. 

\section*{Acknowledgments}

This work was supported by the Schweizerischer Nationalfonds (SNF project 200020\_169120).

\providecommand{\refin}[1]{\\ \textbf{Referenced in:} #1}

\end{document}